\documentclass[a4paper,11pt]{article}
\pdfoutput=1

\usepackage{jheppub}
\usepackage{natbib}
\usepackage{amsmath}
\usepackage{tensor}
\usepackage{multirow}
\usepackage{longtable}
\usepackage[T1]{fontenc} 

\usepackage{acro}

\DeclareAcronym{gqtg}{
    short = {GQTG},
    long = {generalized quasi-topological gravity},
    tag = {acron}
}
\DeclareAcronym{qtg}{
    short = {QTG},
    long = {quasi-topological gravity},
    tag = {acron}
}
\DeclareAcronym{eom}{
    short = {EOM},
    long = {equation of motion},
    tag = {acron}
}
\DeclareAcronym{sqt}{
    short = {ETQT},
    long = {EOM-trivial quasi-topological},
    tag = {acron}
}
\DeclareAcronym{sss}{
    short = {SSS},
    long = {special spherically symmetric},
    tag = {acron}
}
\DeclareAcronym{gss}{
    short = {GSS},
    long = {general spherically symmetric},
    tag = {acron}
}
\DeclareAcronym{kss}{
    short = {KSS},
    long = {Kovtun–Son–Starinets},
    tag = {acron}
}
\newcommand{\dd}{{\mathrm d}}

\title{\boldmath Quasi-topological Gravities on General Spherically Symmetric Metric}

\author[a,b,1]{Feiyu Chen,\note{Corresponding author.}}

\affiliation[a]{Institute of High Energy Physics, Chinese Academy of Sciences, \\Beijing 100049, P.R. China}
\affiliation[b]{School of Physics, University of Chinese Academy of Sciences, \\Beijing 100049, P.R. China}

\emailAdd{chenfy@ihep.ac.cn}

\abstract{In this work we study a more restricted class of quasi-topological gravity theories where the higher curvature terms have no contribution to the equation of motion on general static spherically symmetric metric where $g_{tt} g_{rr} \ne \mathrm{constant}$. We construct such theories up to quintic order in Riemann tensor and observe an important property of these theories: the higher order term in the Lagrangian vanishes identically when evaluated on the most general \textit{non-stationary} spherically symmetric metric ansatz. This not only signals the higher terms could only have non-trivial effects when considering perturbations, but also makes the theories quasi-topological on a much wider range of metrics. As an example of the holographic effects of such theories, we consider a general Einstein-scalar theory and calculate it's holographic shear viscosity.}


\begin{document} 
\maketitle
\flushbottom

\section{Introduction}
\label{sec:intro}
Einstein gravity theory extended with higher order curvature terms plays a relevant role among modified gravity theories. It's predicted by string theory that Einstein gravity should be corrected by an infinite series of higher curvature terms \cite{quartic-heterotic-eft, sigma-ss-eft}. Higher curvature terms have also attracted attentions in holography, they may introduce various new phenomena on the boundary theory. For example, it's shown that including higher curvature terms can lead to the violation of the \ac{kss} shear-viscosity-to-entropy bound $\eta / s \geqslant 1/4\pi$ \cite{kss-bound, holographic-gb, holographic-qtg3}. Holography has also been used to determine the physical bounds of higher curvature couplings by demanding the consistency of the dual CFT \cite{gb-causality, lovelock-causality, holographic-qtg3}.

However, gravity theories with higher curvature terms are generally hard to study. One common way to study a gravity theory is through its black hole solution, however for higher curvature gravities the equations of motion are usually fourth-order differential equations, making analytical solutions hard to come by. It's thus of interest to construct higher curvature theories that admit analytical black hole solutions. On the other hand, the linearized equations of motion of these theories around maximally symmetric spacetimes typically contain fourth derivatives too, so besides the usual massless spin-2 graviton mode, two extra massive modes might appear, the scalar mode and the ghost-like spin-2 mode \cite{general-hogr}. The existence of ghost-like mode signals instability of the AdS vacua and causes unitarity breaking of the dual CFT, it is thus mandatory to decouple (set the mass to infinity) the ghost-like mode when studying holography. The well-known example where these extra modes are absent is Lovelock gravity \cite{lovelock-th, lovelock-2}. Lovelock term of order $k$ vanishes identically when $D \leqslant 2k - 1$ and becomes a total derivative that does not contribute to the equations of motion for $D = 2k$. A total derivative further reduces to a surface term in the action and only contribute topological characteristics, so higher curvature term of this kind is also called \textit{topological} term, no physical effects could emerge when introducing such higher curvature terms.

\Ac{qtg}, on the other hand, is a more intriguing theory in that the equations of motion are drastically simplified when evaluated on some special metric ansatz, and also gives non-trivial contribution at perturbation level. In the broader literature, such theory is defined by that it admits Schwarzschild-like solutions, i.e., the \ac{sss} metric \footnote{Unless otherwise noted, when we say spherically symmetry, we actually mean spherically, planar, or hyperbolic symmetry, corresponding to the curvature of $\dd\Sigma^2$ being positive, zero, or negative, respectively.}
\begin{equation}\label{eq:sss-metric}
    \dd s^2 = -f(r) \dd t^2 + \frac{1}{f(r)} \dd r^2 + r^2\dd\Sigma_{D-2, k}^2
\end{equation}
and the equation of $f(r)$ is algebraic. Cubic quasi-topological gravity was first constructed in \cite{bhs-in-qtg}, it's holographic properties was later studied in \cite{holographic-cubic-qtg}. Higher order ones also exist and they have been studied extensively \cite{new-cubic-5d-qtg, bhs-in-quartic-qtg, quintessential-quartic-qtg, quintic-qtg}. Besides quasi-topological gravities, there's another closely related class of theory worth mentioning, known as the \ac{gqtg}, satisfying \cite{cubic-gqtg, gqtg-at-all-order}
\begin{equation}\label{eq:gqtg-cond1}
    \frac{\delta S_f}{\delta f} = 0, \quad \text{or} \quad \mathcal E_t^t = \mathcal E_r^r
\end{equation}
where $S_f$ denotes the action evaluated on the \ac{sss} metric, and $\mathcal E^{ab} = 1 / \sqrt{|g|} \delta S / \delta g_{ab}$ is the equation of motion. It can be shown that quasi-topological gravity satisfies this condition and thus is a subclass of \ac{gqtg}. Features of \ac{gqtg} have been studied comprehensively at present \cite{einsteinian-cubic-gravity, bhs-in-ecg, 4d-bhs-in-ecg, holographic-ecg, bh-chem-and-holography-in-gqtg, cubic-quartic-gqtg-thermo-holography, higher-gravity-with-mu, gqtg-at-all-order, all-higher-as-gqtg, gqtg-whole-shebang}. In particular, \eqref{eq:gqtg-cond1} implies \cite{all-higher-as-gqtg} the equation of $f(r)$ is at most second order, the existence of Schwarzschild-like solutions, and most importantly, the decoupling of the extra massive modes.

We are more interested in quasi-topological gravities whose higher curvature terms do not contribute to the equation of motion when evaluated on some special metric ansatz. We refer to such class of theories as \ac{sqt} gravities, meaning they have trivial \acf{eom} contributions. These theories obviously satisfy \eqref{eq:gqtg-cond1} and thus ghost-free. They are first considered in \cite{lyz1} for Ricci polynomials, where \ac{sqt} terms were constructed up to tenth order in Ricci tensor both on the \ac{sss} metric and \ac{gss} metric
\begin{equation}\label{eq:gss-metric}
    \dd s^2 = -h(r) \dd t^2 + \frac{1}{f(r)} \dd r^2 + r^2\dd\Sigma_{D-2, k}^2
\end{equation}
The advantage of this definition of quasi-topological gravity is that black hole solutions in the original gravity theory in the form of \eqref{eq:sss-metric} or \eqref{eq:gss-metric} simply continue to be solutions when the corresponding quasi-topological terms are introduced. This could be relevant when matter are included, where even with the equation of $f(r)$ being algebraic, the inclusion of matter terms could make the system unintegratable. In this work we are specifically interested in \ac{sqt} gravities on \ac{gss} metric \eqref{eq:gss-metric}, but not just limited to Ricci gravities. Such metric is the most general ansatz for spacetimes with spherical/planar/hyperbolic symmetry, thus could include a wider range of solutions. An important class of \ac{gss} metric is black hole with scalar hair. Hairy solutions typically have a rich phase structure and in holography they may be used to describe superconductors \cite{building-holographic-superconductor, holographic-superconductor}. It was shown that even Einstein gravity with a minimally coupled self-interacting scalar field could result in a hairy solution \cite{mtz-bh}. It's thus interesting to investigate the effects of including higher curvature terms in these solutions. There has been works on the hairy solutions with conformally coupled scalar in higher curvature gravities \cite{hairy-bh-in-qtg, quadratic-hairy-bh}.

In this work we focus on \ac{sqt} gravities on \ac{gss} metric\footnote{In the rest of the paper we'll simply refer to \ac{sqt} gravity on \ac{gss} metric as \ac{sqt} gravity, unless noted explicitly.} and construct such quasi-topological terms up to quintic order in Riemann tensor, including dimension-generic ones and dimension-specific ones. For the latter only $D \geqslant 3$ is considered since at $D = 2$ the Riemann tensor has only one non-zero component. We also find it possible to construct dimension-independent combinations at quartic and quintic order. We then notice that these theories satisfy a much stronger condition: the \ac{sqt} terms vanish identically when evaluated on the non-stationary spherically symmetric metric! That is, metric of the following form
\begin{equation}\label{eq:ggss-metric}
    \dd s^2 = -h(t, r)\dd t^2 + 2b(t, r)\dd t\dd r + \frac{1}{f(t, r)}\dd r^2 + r^2\dd\Sigma_{D-2,k}^2
\end{equation}
On the one hand, this further restricts the effects \ac{sqt} term could possibly have, such as no thermodynamics contribution. This possibly simplifies the problem since introducing \ac{sqt} terms won't lead to any new phase transitions. Thus one may only seek for non-trivial effects of \ac{sqt} terms by considering perturbations. On the other hand, this indicates that \ac{sqt} terms are also quasi-topological on a much wider range kinds of metrics, e.g., Friedmann-Roberson-Walker metric, making it simpler to study, e.g., the effects of \ac{sqt} terms on cosmic perturbations.

This paper is organized as follows. In section \ref{sec:theory-construct} we construct explicitly the \acl{sqt} gravity theories, up to quintic order. In section \ref{sec:props} we discuss the basic properties of the obtained theories, mainly the implications of the vanishing of \ac{sqt} terms evaluated on the metric \eqref{eq:ggss-metric}. As an example to study the physical effects of \ac{sqt} terms, in section \ref{sec:sv} we consider a general Einstein-scalar theory extended with \ac{sqt} terms and calculate its holographic shear viscosity.

\section{Construction of the theory}
\label{sec:theory-construct}

For a general Lagrangian constructed from metric and Riemann tensor $\mathcal L(g_{ab}, R_{abcd})$ the equation of motion can be written as \cite{lanczos-lovelock}
\begin{equation}
    \mathcal E^{ab}[\mathcal L] = \frac{1}{\sqrt{|g|}}\frac{\delta S}{\delta g_{ab}} = P\indices{^a_c_d_e} R^{bcde} - \frac12 g^{ab}\mathcal L + 2\nabla_c\nabla_d P^{acdb},
    \quad P^{abcd} = \frac{\partial\mathcal L}{\partial R_{abcd}}
\end{equation}
where $S = \int\dd^D x\sqrt{|g|}\mathcal L$ is the action. An \ac{sqt} term $\mathcal Q$ in the Lagrangian satisfies that it does not contribute to the equation of motion when evaluated on \eqref{eq:gss-metric}, namely
\begin{equation}\label{eq:qtg-cond}
    \left.\mathcal E^{tt}[\mathcal Q]\right|_{h, f} = \left.\mathcal E^{rr}[\mathcal Q]\right|_{h, f} = 0
\end{equation}
which is equivalent to
\begin{equation}\label{eq:qtg-cond-alt}
    \frac{\delta}{\delta h}\int\dd^D x\left.\sqrt{|g|}\mathcal Q\right|_{h, f} = 
    \frac{\delta}{\delta f}\int\dd^D x\left.\sqrt{|g|}\mathcal Q\right|_{h, f} = 0
\end{equation}
where $\dots |_{h, f}$ denotes $\dots$ evaluated on the metric \eqref{eq:gss-metric}. At a given order, we first write down the most general Riemann polynomial of that order with undetermined coefficients and substitute the ansatz \eqref{eq:gss-metric} into it. The non-zero Riemann tensor components of \eqref{eq:gss-metric} are
\begin{align*}
    R^{\hat t\hat r}{}_{\hat t\hat r} & = \frac{f(r) h'^2(r) - h \left[f'(r) h'(r) + 2 f(r) h''(r)\right]}{4 h^2(r)} \\
    R^{\hat ti}{}_{\hat tj} & = -\delta^i_j \frac{f(r) h'(r)}{2 r h(r)} \\
    R^{\hat ri}{}_{\hat rj} & = -\delta^i_j \frac{f'(r)}{2r} \\
    R^{ij}{}_{kl} & = (\delta^i_k \delta^j_l - \delta^i_l \delta^j_k) \frac{k - f(r)}{r^2}
\end{align*}
where $ijkl$ are indices of the $(D - 2)$ dimension subspace, and equivalent components are not shown. By varying and integrating by parts with respect to $h(r)$ and $f(r)$ we get two algebraic equations containing the undetermined coefficients, we then further convert them into a linear system about the undetermined coefficients by regarding them as polynomials in $r, h(r), f(r)$ and their derivatives and requiring all coefficients vanish. The solution space is given by the null space of the resulting linear system. For dimension-generic solutions, we take null space directly, and for dimension-specific ones we substitute the dimension first and then take null space, since there may be more linear independent solutions at lower dimensions.

\subsection{Cubic order}
There are 8 Riemann scalars at cubic order and the most general cubic Riemann polynomial is given by their linear combination
\begin{align}
    \mathcal Q^{(3)} & = 
    e_1 R\indices{_a^b_c^d} R\indices{_b^e_d^f} R\indices{_e^a_f^b}
    + e_2 R\indices{_a_b^c^d} R\indices{_c_d^e^f} R\indices{_e_f^a^b}
    + e_3 R_{abcd}R\indices{^a^b^c_e} R^{de}
    + e_4 R_{abcd} R^{abcd} R\nonumber\\
    & \qquad + e_5 R_{abcd} R^{ac} R^{bd}
    + e_6 R_a^b R_b^c R_c^a
    + e_7 R_a^b R_b^a R
    + e_8 R^3
\end{align}
We found only one dimension-generic solution in this case, the coefficients $e_i$ are given by
\begin{align}\label{eq:cubic-qtg-solution}
    & e_1 = 22 - 26 D + 9 D^2 - D^3, e_2 = \frac{3 D^2}{4}-\frac{15 D}{4} + 4, e_3 = -3 (D-3) (D-1)\nonumber\\
    & e_4 = \frac{3 (D-3)}{2}, e_5 = 3 \left(D^2-5 D+8\right), e_6 = 6 D-14, e_7 = 3 - 3D, e_8 = 1
\end{align}
As mentioned earlier, \ac{sqt} gravity is a subclass of generalized quasi-topological gravity, so the solution \eqref{eq:cubic-qtg-solution} must be a special case of cubic generalized quasi-topological gravity. In fact, by setting
\begin{equation*}
    c_1 = 22 - 26 D + 9 D^2 - D^3, c_2 = \frac{3 D^2}{4}-\frac{15 D}{4} + 4, c_3 = -3 (D-3) (D-1)
\end{equation*}
in (2.6) of \cite{cubic-gqtg} we get our solution \eqref{eq:cubic-qtg-solution}. As there's only one solution, it's not possible to construct dimension independent solution at cubic order.

Now we turn to dimension-specific solutions. First we found that for $D > 6$ the number of independent solutions is always one, meaning they are covered by the dimension-generic solution, so we only need to consider $3 \leqslant D \leqslant 6$. We get two linear independent solutions at $D=4$ and $D=6$ respectively, five solutions at $D=3$ and one solution again at $D=5$. The solutions are given in table \ref{tab:cubic-qtg-dimspec-sols}.
\begin{table}[t!]
    \centering
    \begin{tabular}{|c|c|}
        \hline
        $D$ & $\{e^i\}$ \\
        \hline
        \multirow{5}{*}{3}
        & $(-8,5,-12,0,0,0,0,1)$ \\
        & $(-2,\frac{3}{2},-4,0,0,0,1,0)$ \\
        & $(0,\frac{1}{2},-\frac{3}{2},0,0,1,0,0)$ \\
        & $(-1, \frac14, -1, 0, 1, 0, 0, 0)$ \\
        & $(0, 1, -4, 1, 0, 0, 0, 0)$ \\
        \hline
        \multirow{2}{*}{4}
        & $(16,-8,36,-3,-24,-8,0,1)$ \\
        & $(2,-1,5,-\frac{1}{2},-4,-2,1,0)$ \\
        \hline
        5 & $(-8,4,-24,3,24,16,-12,1)$ \\
        \hline
        \multirow{2}{*}{6}
        & $(64,-14,60,-3,-48,-8,0,1)$ \\
        & $(6,-\frac{3}{2},7,-\frac{1}{2},-6,-2,1,0)$\\
        \hline
    \end{tabular}
    \caption{Dimension-specific solutions of cubic \ac{sqt} gravity}
    \label{tab:cubic-qtg-dimspec-sols}
\end{table}
However, not all solutions are non-trivial. It could happen that some solutions vanish identically on any metric, just like Lovelock terms in $D < 2n$, this is possible for dimension-specific cases.  Firstly we note that all the solutions in table \ref{tab:cubic-qtg-dimspec-sols} have included the cubic Lovelock term, especially the five dimensional solution which simply coincides with it. So we are left with 4 solution for $D = 3$, one solution for $D = 4$ and $D = 6$ respectively.

Besides Lovelock terms themselves, another kind of combinations that vanish in lower dimensions may be constructed from their equation of motion. For example, in $D \leqslant 4$, the 4D Lovelock, or Gauss-Bonnet term
\begin{equation}
    \mathcal X^{(4)} = R^2 - 4R_{ab}R^{ab} + R_{abcd}R^{abcd}
\end{equation}
is topological, so its equation of motion contribution should vanish
\begin{align}
    \mathcal E^{ab}[\mathcal X^{(4)}] & = \frac{1}{\sqrt{|g|}}\frac{\delta}{\delta g_{ab}}\int\dd^D x\sqrt{|g|}\mathcal X^{(4)}\nonumber\\
    & = -4 R^{ac} R^b_c + 2 R^{ab}R - 4R^{cd}R\indices{^a_c^b_d} + 2R^{acde} R\indices{^b_c_d_e} + \frac12 g^{ab} \mathcal X^{(4)} = 0
\end{align}
we can thus construct another vanishing Riemann polynomial
\begin{align}\label{eq:4dvanish1}
    \mathcal E^{ab}[\mathcal X^{(4)}] R_{ab} & = -4R^a_b R^b_c R^c_a + 4R_{ab}R^{ab}R - \frac12 R^3 - 4R^{ab} R^{cd}R_{acbd}\nonumber\\
    & \qquad - \frac12 R^{abcd}R_{abcd}R + 2R^{ab}R\indices{_a^c^d^e}R_{bcde} = 0
\end{align}
It can be shown that the space spanned by \eqref{eq:4dvanish1} and 4D Lovelock term is isomorphic to the $D = 4$ solution in table \ref{tab:cubic-qtg-dimspec-sols}, thus both solutions are trivial. In three dimensions, the Gauss-Bonnet term vanishes, there are three more vanishing Riemann polynomials \footnote{$\mathcal L = 0$ implies $\partial\mathcal L / \partial R^{abcd} = 0$ if the identity $\partial\mathcal L / \partial g_{mn} = (\partial\mathcal L / \partial R^{abcd}) (\partial R^{abcd} / \partial g_{mn})$ gives no less equations than the independent components of $\partial\mathcal L / \partial R^{abcd}$, which is true for $D \leqslant 3$.}
\begin{equation}\label{eq:3dvanishing-terms}
    R \mathcal X^{(4)}, \qquad \frac{\partial\mathcal X^{(4)}}{\partial R^{abcd}} R^{ac} R^{bd},
    \qquad \frac{\partial\mathcal X^{(4)}}{\partial R^{abcd}} R^{cdef} R\indices{_e_f^a^b}
\end{equation}
Again, the space spanned by these three terms, \eqref{eq:4dvanish1} and 4D Lovelock, is isomorphic to the $D = 3$ solution in table \ref{tab:cubic-qtg-dimspec-sols}, which means they are all trivial. This result is also consistent with \cite{aspects-of-3d-higher-grav}, which states non-trivial \ac{gqtg} of order less than six in curvature does not exist in $D = 3$.

In summary, only one (linear combination) of the $D = 6$ solution in table \ref{tab:cubic-qtg-dimspec-sols} is non-trivial. This 6D solution is also covered by the dimension-generic solution \eqref{eq:cubic-qtg-solution}. We finally conclude that cubic \ac{sqt} term is completely given by \eqref{eq:cubic-qtg-solution} and it's only non-trivial for $D \geqslant 6$.

Before moving on to higher orders, it's worthy to compare the definition of \ac{sqt} gravity on \ac{gss} metric \eqref{eq:qtg-cond} with \ac{sqt} gravity on \ac{sss} metric, that is
\begin{equation}\label{eq:qtg-sss-cond}
    \left.\mathcal E^{tt}[\mathcal Q^{\mathrm{SSS}}]\right|_{f, f} = \left.\mathcal E^{rr}[\mathcal Q^{\mathrm{SSS}}]\right|_{f, f} = 0
\end{equation}
Such class of gravities is already known, a $D = 4$ example is the density $\mathcal C$ given in (2.16) of \cite{cubic-gqtg}. One can expect the condition \eqref{eq:qtg-sss-cond} is weaker than ours \eqref{eq:qtg-cond}. In fact, after repeating the above computation on cubic order using this condition, we find the solutions are almost the same, the only difference is an extra non-trivial solution at $D = 4$
\begin{equation}
    \mathcal Q^{\mathrm{SSS}, D = 4} = \frac{1}{2}R_a^b R_b^a R - 2 R^{ac}R^{bd}R_{abcd} - \frac{1}{4}RR_{abcd}R^{abcd} + R^{de}R_{abcd}R^{abc}{}_e
\end{equation}
which is exactly the density $\mathcal C$, and it's easy to check that it gives non-vanishing equation of motion contribution when evaluated on the \ac{gss} metric \eqref{eq:gss-metric}.

\subsection{Higher orders}
The method of solving higher order \ac{sqt} terms is exactly the same as we used in the cubic case. The only difficulty is enumerating all possible Riemann scalars, as the number of independent Riemann scalars grows rapidly in higher orders. We get 26 scalars at quartic order \cite{fulling-tensor-polynomials} and 85 scalars at quintic order.

For quartic case we get 12 linear independent dimension-generic solutions, and at $D = 3$, $D = 4$ and $D = 8$ we get 22, 15 and 13 solutions respectively, all other dimensions have the same number of solutions as dimension-generic case. Again, the extra solution at $D = 8$ is the 8D Lovelock term. Using the method explained in appendix \ref{sec:apdx-quartic-order}, we found that the $D = 3$ solutions are all trivial, making it in consistency with \cite{aspects-of-3d-higher-grav}, this is also true for the quintic case below. The explicit list of solutions is lengthy and given in appendix. Remarkably, we also found 3 dimension-independent solutions
\begin{align}
    \mathcal Q^{(4), *, 1} & = R^{abcd}\left(R\indices{_a^e_c^f}R\indices{_b^g_e^h}R_{dgfh}
    - R\indices{_a_b^e^f}R\indices{_c^g_e^h}R_{dgfh}
    - \frac14 R\indices{_a_b_c^e}R\indices{_d^f^g^h}R_{efgh}\right)\\
    \mathcal Q^{(4), *, 2} & = R\indices{_a_b^e^f}R^{abcd}\left(R\indices{_c_e^g^h}R_{dfgh} - \frac12 R\indices{_c_d^g^h}R_{efgh}\right) \\
    \mathcal Q^{(4), *, 3} & = R^{ab}R^{cd}\left(R\indices{_a^e_c^f}R_{bedf} - \frac12 R\indices{_a_c^e^f}R_{bdef}\right) - \frac12 R^{ab}R_a^c R\indices{_b^d^e^f}R_{cdef}
\end{align}
The situation is similar for quintic case, we get 61 dimension-generic solutions and 80, 67, 62 dimension-specific solutions at $D = 3$, $D = 4$, $D = 10$ respectively, for dimension-independent case we get 29 solutions. However as the solutions of the quintic case are too lengthy, we only present some representative solutions in the appendix and the full solution set can be found in the supplementary material.

\section{Properties and discussions}
\label{sec:props}
Having constructed the desired theories we now move on to their physical effects. The first property we noticed is the \ac{sqt} term vanishes when evaluated on the metric \eqref{eq:gss-metric}
\begin{equation}\label{eq:q-vanish-on-gss}
    \left.\mathcal Q^{(n)}\right|_{h,f} = 0
\end{equation}
The free energy can be obtained by evaluating the Euclidean action with compactified time direction. Since our metric is static, the Euclidean action only differs from the Minkowski action by a minus sign, \eqref{eq:q-vanish-on-gss} implies the vanishing of the free energy contribution from \ac{sqt} term, which then further implies the entropy and thermodynamic energy contribution should also vanish, thus \ac{sqt} term completely has no thermodynamics effects.

To verify the consistency of the above conclusion we need to evaluate the Wald entropy and thermodynamic energy. The Wald entropy is given by \cite{wald-entropy}
\begin{equation}\label{eq:wald-entropy}
    S_{\mathrm{Wald}} = -2\pi\oint P^{abcd}\epsilon_{ab}\epsilon_{cd} \, \dd\Sigma,
    \qquad P^{abcd} = \frac{\partial\mathcal L}{\partial R_{abcd}}
\end{equation}
where the integration is taken at the horizon, $\epsilon_{ab}$ is the binormal to the horizon, $\dd\Sigma$ is the volume form of the horizon surface. Using the method similar to \cite{gqtg-at-all-order} one can show that $\mathcal Q^{(n)}|_{h,f} = 0$ implies
\begin{equation}\label{eq:vanish-of-q-pd-r}
    \left.\frac{\partial Q^{(n)}}{\partial R_{abcd}}\right|_{h,f} = 0
\end{equation}
thus the $\mathcal Q^{(n)}$ contribution to \eqref{eq:wald-entropy} must also vanish. It remains to calculate the energy, which can be done holographically by calculating the $tt$ component of the boundary stress tensor $T^{tt} = (2/\sqrt{|h|}) \delta S / \delta h_{tt}$ where $h_{ab}$ is the boundary metric. The surface term and counter term also need to be taken into account, but since $\mathcal Q^{(n)}$ vanishes, no new diverges appear, the counter term contribution is zero. The surface term can be constructed by introducing an auxiliary field $\Phi_{ab} = P_{acbd}n^c n^d$ \cite{f-riemann}
\begin{equation}
    S_\partial = \frac{1}{8\pi}\oint_{\partial M}\dd^{D - 1} x \sqrt{|h|} \Phi^{ab}K_{ab}
\end{equation}
where $n^a$ is the normal vector of the boundary, $K_{ab} = \nabla_a n_b$ is the exterior curvature and $h_{ab} = g_{ab} - n_a n_b$. Note that when varying this term, $\Phi^{ab}$ should be kept fixed. So we immediately see from \eqref{eq:vanish-of-q-pd-r} that the surface term contribution to $T^{tt}$ should vanish. Furthermore, because $\mathcal Q^{(n)}$ vanishes on \eqref{eq:gss-metric}, it's invariant under the variation $h(r) \to h(r) + \delta h(r)$, so we have $\delta\sqrt{|g|}\mathcal Q^{(n)} / \delta h_{tt} = 0$. We thus conclude the energy obtained via holography also vanishes.

As mentioned earlier, we actually found a much stronger conclusion than \eqref{eq:q-vanish-on-gss}, that is $\mathcal Q^{(n)}$ also vanishes when evaluated on the general non-stationary spherically symmetric metric \eqref{eq:ggss-metric}. It's straightforward to evaluate a given \ac{sqt} terms on \eqref{eq:ggss-metric} and check that it vanishes. In practice, the check was done using an equivalent metric
\begin{equation}\label{eq:ggss-metric-alt1}
    \dd s^2 = -h(t, r)\dd t^2 + 2b(t, r)\dd t\dd r + \left[\frac{1}{f(t, r)} - \frac{b^2(t, r)}{h(t, r)}\right]\dd r^2 + r^2\dd\Sigma_{D-2, k}^2
\end{equation}
The advantage of it is the components of the inverse metric contain no fraction, reducing the computation cost. The check was done for all cubic and quartic \ac{sqt} terms, but at quintic order we encountered extreme computation difficulties so we ended up only checked the solutions listed in \eqref{eq:quintic-qtgs}. We could then conjecture that the condition for \ac{sqt} terms \eqref{eq:qtg-cond-alt} implies that they vanish when evaluated on \eqref{eq:ggss-metric}.

The vanishing of $\mathcal Q^{(n)}$ on \eqref{eq:ggss-metric} makes it quasi-topological on a much wider range of metrics, e.g. the FRW metric
\begin{equation}
    \dd s^2 = -\dd t^2 + a^2(t)\left[\frac{\dd r^2}{1 - kr^2} + r^2\dd\Omega^2\right]
\end{equation}
after defining a new radial coordinate $r'$ by $r' = r a(t)$ we get
\begin{equation}
    \dd s^2 = \frac{kr'^2 + r'^2\dot a - a^2}{a^2 - kr'^2}\dd t^2 - 2\frac{r' a \dot a}{a^2 - k r'^2}\dd t \dd r' + \frac{a^2}{a^2 - k r'^2}\dd r'^2 + r'^2 \dd\Omega^2
\end{equation}
which has the same form as \eqref{eq:ggss-metric}. Another example implies $\mathcal Q^{(n)}$ has no a-charge contribution. In $2n$ dimensional CFTs, the central a, c charges appears as coefficients in the trace of the stress tensor \cite{weyl-anomaly}
\begin{equation}
    \left\langle T^\mu_\mu\right\rangle \sim -a \mathcal X^{(2n)} + \sum_i c_i I_i^{(2n)}
\end{equation}
where $\mathcal X^{(2n)}$ is the $2n$ dimensional Lovelock term, $I_i^{(2n)}$ are conformal invariants in $2n$ dimensional space. Generally the central charges can be calculated holographically by evaluating the action on the the FG expansion metric and identifying the $\rho^{-1}$ term as the trace anomaly \cite{fg-expansion-weyl}, but to solely extract the a-charge one may use a specific metric with conformally flat boundary, e.g., an $S^{2n}$ \cite{lyz2}
\begin{equation}\label{eq:reduced-fg-ex}
    \dd s^2 = \frac{L^2}{4\rho^2}\dd\rho^2 + \frac{f(\rho)}{\rho}\left(\frac{\dd r^2}{1 - r^2} + r^2 \dd\Omega_{D - 2}^2\right)
\end{equation}
again, by redefining $r \to r \sqrt{\rho / f(\rho)}$ this metric can be put into the form of \eqref{eq:ggss-metric}, thus $\mathcal Q^{(n)}$ also vanish on \eqref{eq:reduced-fg-ex}, it doesn't contribute to the a-charge.

The vanishing of $\mathcal Q^{(n)}$ on the more general metric \eqref{eq:ggss-metric} largely reduces the possible effects it could have when introduced to some gravity theory. To seek for non-trivial effects one may only consider the perturbations of it around the metric \eqref{eq:gss-metric}, which in holography includes shear viscosity and heat current, corresponding to perturbations $h_{x_1 x_2}$ and $h_{tx_1}$ respectively. In the next section we'll consider the holographic shear viscosity as an example to study.

\section{Holographic shear viscosity}
\label{sec:sv}
We consider a general Einstein-scalar theory with the Lagrangian
\begin{equation}\label{eq:es-lag}
    \mathcal L_{\mathrm{ES}} = \frac{1}{16\pi}\left[R - \frac12 \nabla_a\phi\nabla^a\phi - V(\phi)\right]
\end{equation}
we are interested in the non-extremal \footnote{The extremal limit $T \to 0$ and hydrodynamic limit $\omega \to 0$ generally don't commute, which will complicate the discussion.} asymptotic AdS black hole solutions of \eqref{eq:es-lag}, so we consider the following planar black hole ansatz
\begin{equation}\label{eq:planar-bh-metric}
    \dd s^2 = -f(r)e^{-\sigma(r)}\dd t^2 + \frac{1}{f(r)}\dd r^2 + U(r)\dd\vec x_{D - 2}^2
\end{equation}
Note that there's one gauge freedom in the three functions $f(r)$, $\sigma(r)$, $U(r)$. Near the boundary we have $f(r \to \infty) = U(r \to \infty) = r^2 / L^2$, $\sigma(r \to \infty) = 0$, where $L$ is the AdS radius. The horizon is at $r = r_h$, satisfies $f(r_h) = 0$.

The reason for considering a hairy black hole solution instead of a simpler Schwarzschild solution is that we think the shear viscosity of hairy solution may have some unique features comparing to Schwarzschild solution. Also, as the metric of hairy solution is \ac{gss} metric but not \ac{sss} metric, and \ac{sqt} term has vanishing equation of motion contribution on \ac{gss} metric, we think it's worthy to consider a hairy solution. Besides, we can easily go back to Schwarzschild case by substituting
\begin{equation}
    f(r) = \frac{r^2}{L^2}\left[1 - \left(\frac{r_h}{r}\right)^2\right],
    \quad \sigma(r) = 0,
    \quad U(r) = \frac{r^2}{L^2},
    \quad \phi(r) = 0
\end{equation}
into the result.

The temperature and entropy density are respectively given by
\begin{equation}
    T = \frac{1}{4\pi}f'(r_h)e^{-\sigma(r_h)/2},
    \qquad s = \frac14 U^{D / 2 - 1}(r_h)
\end{equation}
Assuming $\phi$ only depends on $r$, the equation of motion gives
\begin{align}\label{eq:es-eoms}
    (D-2) f U' \phi' + U \left(2 f' \phi' - f \sigma' \phi' + 2f \phi'' - 2 V'\right) & = 0\nonumber\\
    \left(D^2-7 D + 10\right) f U'^2 + 2 (D - 2) U \left(f' U' + 2f U''\right) + 2 U^2 \left(f \phi'^2 + 2 V\right) & = 0\nonumber\\
    2 U \left[(D - 2) U''(r)+U(r) \phi'^2\right] + (D - 2) U U' \sigma' - (D - 2) U'^2 & = 0\nonumber\\
    U\left(-(D-4) f' U' + f\left((D - 3) \sigma' U' + 2 U''\right)\right)+(D - 4) f U'^2 &\nonumber\\
    + U^2 \left(-2 f''+3 f' \sigma ' - f \left(\sigma'^2 - 2\sigma''\right)\right) & = 0
\end{align}
The last equation can be integrated to give a radially conserved quantity, as in \cite{hairy-thermodynamics}
\begin{equation}\label{eq:conserved-q}
    Q = e^{-\sigma / 2} U^{D/2 - 2} \left[\left(f' - f\sigma'\right)U - f U'\right]
\end{equation}
Evaluating it at the horizon gives
\begin{equation}
    Q = e^{-\sigma(r_h) / 2}U^{D/2 - 1}(r_h)f'(r_h) = 16\pi T s
\end{equation}
To calculate the holographic shear viscosity we employ the pole method as proposed in \cite{pole-method}. Define a new radial coordinate $z$ by $r = r_h / (1 - z)$, \eqref{eq:planar-bh-metric} becomes
\begin{equation}\label{eq:planar-metric-2}
    \dd s^2 = \frac{r_h^2}{(1 - z)^4} \frac{1}{f(\frac{r_h}{1 - z})} \dd z^2 - f\left(\frac{r_h}{1 - z}\right)\exp\left[-\sigma\left(\frac{r_h}{1 - z}\right)\right] \dd t^2 + U\left(\frac{r_h}{1 - z}\right) \dd\vec x_{D - 2}^2
\end{equation}
Now add perturbation to \eqref{eq:planar-metric-2} by shifting the basis $\dd x^1 \to \dd x^1 + \varepsilon e^{-i\omega t}\dd x^2$, substitute the resulting metric into the Lagrangian and expand it to quadratic order in $\varepsilon$. Note that the perturbation should be kept second order in the metric, and since the perturbation only involves spatial components, the matter sector of the Lagrangian \eqref{eq:es-lag} has no contribution. The shear viscosity can be calculated from the residue of the Lagrangian at $z = 0$
\begin{equation}\label{eq:pole-formula}
    \eta = -8\pi T \lim_{\omega, \varepsilon \to 0} \frac{\mathrm{Res}_{z = 0} \sqrt{|g|}\mathcal L}{\varepsilon^2 \omega^2}
\end{equation}
Note that the above expression for the shear viscosity is linear in $\mathcal L$ so we can compute the contribution to the shear viscosity of different terms in the Lagrangian separately, but keep in mind only the summed result has physical meaning. For the Einstein-scalar theory in \eqref{eq:es-lag}, the contribution is given by
\begin{equation}
    \eta^{(0)} = \frac{1}{16\pi}U^{D/2 - 1}(r_h)
\end{equation}
which results in the standard shear-viscosity-to-entropy ratio $\eta^{(0)} / s = 1/4\pi$, i.e., the existence of the scalar hair have no effect on the shear viscosity.

Next we introduce \ac{sqt} terms to \eqref{eq:es-lag} by defining the new Lagrangian as $\mathcal L' = \mathcal L_{\mathrm{ES}} + (\lambda / 16\pi)\mathcal Q^{(n)}$, the equations of motion aren't altered by $\mathcal Q^{(n)}$ and it's straightforward to evaluate \eqref{eq:planar-metric-2} on them and then apply \eqref{eq:pole-formula} to obtain the shear viscosity, the results are expressed in $f(r)$, $\sigma(r)$, $U(r)$ and their derivatives at $r = r_h$. Interestingly, by making use of the radially conserved quantity \eqref{eq:conserved-q} we are only left with $\sigma(r_h)$, $\sigma'(r_h)$, $U(r_h)$ and its derivatives. The contribution from $\mathcal Q^{(3)}$ is given by\footnote{We use $\hat\eta$ to denote a ``partial'' shear viscosity.}
\begin{align}
    \hat\eta_{\mathcal Q^{(3)}} & = \frac{\lambda}{16\pi}\frac{3}{16}(D-5) (D-4) (D-2)^2 Q^2 e^{\sigma (r_h)} U^{-D/2-1}(r_h)\nonumber\\
    & \qquad \left[(D+2) U'^2(r_h)-2 U(r_h) U''(r_h)-U(r_h) \sigma'(r_h) U'(r_h)\right]
\end{align}
Notice that this result is only non-zero for $D \geqslant 6$, otherwise the \ac{sqt} term is trivial, as discussed earlier. The shear-viscosity-to-entropy ratio in this case is given by
\begin{equation}\label{eq:cubic-qtg-eta-to-s}
    \frac{\eta_{\mathcal Q^{(3)}}}{s} = \frac{1}{4\pi}\left\{1 + \frac{3\lambda}{16}(D-5) (D-4) (D-2)^2 Q^2 e^{\sigma}\left[
        (D + 2)\frac{U'^2}{U^2} - 2\frac{U''}{U} - \sigma'\frac{U'}{U}
    \right]\right\}
\end{equation}
where all functions are evaluated at the horizon. In Schwarzschild solution this reduces to
\begin{equation}
    \left.\frac{\eta_{\mathcal Q^{(3)}}}{s}\right|_{\mathrm{Schwarzschild}} = \frac{1}{4\pi}\left\{1 + \frac{3\lambda}{4L^4} (D - 5)(D - 4)(D - 2)^2 (D - 1)^2 (D + 1)\right\}
\end{equation}
This indicates a possible violation of the \ac{kss} bound. This is expected, since there's already a lot of examples of higher curvature theory that violates \ac{kss} bound. Indeed, to confirm that \eqref{eq:cubic-qtg-solution} violates the bound requires determining the physical bound of the coupling constant $\lambda$, we will not discuss it here.

For quartic and quintic case, we found that the $D = 4$ solutions are all analytical at $z = 0$ when evaluated on the metric \eqref{eq:planar-metric-2} and thus does not contribute to the shear viscosity. For the dimension-generic solutions, we found their shear viscosity contribution can written in the form
\begin{equation}\label{eq:sv-contrib-form}
    \hat\eta_{\mathcal Q^{(n)}} = \frac{\lambda}{16\pi}e^{\frac{n - 1}{2}\sigma} Q^{n-1}U^{-1 - \frac{n-2}{2}D} U'^{n - 3}\left(a U'^2 + b UU'\sigma' + 2b UU''\right)
\end{equation}
which in Schwarzschild solution reduces to
\begin{equation}
    \left.\hat\eta_{\mathcal Q^{(n)}}\right|_{\mathrm{Schwarzschild}} = \frac{\lambda}{16\pi}2^{n-1} (a + b) (D - 1)^{n-1} r_h^{D - 2} L^{-D - 2n + 4}
\end{equation}
For a specific \ac{sqt} term, the coefficients $a$, $b$ only depends on the dimension $D$, their explicit values are given in the appendix. 

\section{Conclusions}
\label{sec:conc}

\Ac{sqt} gravities can be thought as a class of higher curvature gravity theories whose higher curvature terms give no contribution to the equations of motion when evaluated on the metric \eqref{eq:gss-metric}, but could have non-trivial perturbations around it. In this case black hole solutions of the corresponding Einstein gravity continues to be solution when the higher curvature terms are included, making it much easier to study its higher curvature effects. In this work we constructed such theory up to quintic order in the Riemann tensor. Most remarkably, we found that all \ac{sqt} terms we constructed actually vanish when evaluated on the most general non-stationary spherically symmetric metric \eqref{eq:ggss-metric}. On the one hand, this makes these terms have no contribution on the thermodynamics and holographic a-charge. More importantly, on the other hand, this makes them quasi-topological on a much wider kinds of metrics, e.g., the FRW metric and the Vaidya metric. This opens a large gate of possible applications of such \ac{sqt} gravity theories, such as one could study the effects of these terms on the cosmic perturbations.

As an example to study the non-trivial effects of the \ac{sqt} terms we calculated the holographic shear viscosity of a general Einstein-scalar theory. The results can be put into a simple form \eqref{eq:sv-contrib-form}. As expected, the \ac{kss} bound could possibly be violated due to the nature of higher curvature gravities.

\appendix
\allowdisplaybreaks
\section{Quartic and quintic \ac{sqt} gravities}
In this section we list all solutions of quartic and quintic \ac{sqt} terms. The full set of solutions is also available in the supplementary material, in the form of Mathematica \texttt{.wl} file, with further instructions included in the usage messages.

\subsection{Quartic order}
\label{sec:apdx-quartic-order}

For quartic order, the most general Riemann polynomial is
\begin{align}
    \mathcal Q^{(4)} & = 
        e_1 R^4 
        + e_2 R^2 R^{ab}R_{ab} 
        + e_3 RR^a_bR^b_c R^c_a
        + e_4 (R^{ab}R_{ab})^2
        + e_5 R^a_b R^b_c R^c_d R^d_a
        \nonumber\\& \quad + e_6 RR^{ac}R^{bd}R_{abcd}
        + e_7 R^{ac}R^b_e R^{ed} R_{abcd}
        + e_8 R^2 R^{abcd}R_{abcd}
        + e_9 RR^{de}R\indices{^a^b^c_d}R_{abce}
        \nonumber\\& \quad + e_{10} R^{ab}R_{ab}R^{cdef}R_{cdef}
        + e_{11} R^{ab}R_a^c R\indices{^d^e^f_b}R_{defc}
        + e_{12} R^{ab}R^{cd}R\indices{^e^f_a_c}R_{efbd}
        \nonumber\\& \quad + e_{13} R^{ab}R^{cd}R\indices{^e_a^f_b}R_{ecfe}
        + e_{14} R^{ab}R^{cd}R\indices{^e_a^f_c}R_{ebfd}
        + e_{15} RR^{abcd}R\indices{_a_b^e^f}R_{efcd}
        \nonumber\\& \quad + e_{16} RR^{abcd}R\indices{_a^e_c^f}R_{bedf}
        + e_{17} R^{ab}R\indices{_a^c_b^d}R\indices{^e^f^g_c}R_{efgd}
        + e_{18} R^{ab}R^{cdef}R\indices{_c_d^g_a}R_{efgb}
        \nonumber\\& \quad + e_{19} R^{ab}R^{cdef}R\indices{_c^g_e_a}R_{dgfb}
        + e_{20} (R^{abcd}R_{abcd})^2
        + e_{21} R^{abcd}R\indices{_a_b_c^e}R\indices{^f^g^h_d}R_{fghe}
        \nonumber\\& \quad + e_{22} R^{abcd}R\indices{_a_b^e^f}R\indices{_e_f^g^h}R_{cdgh}
        + e_{23} R^{abcd}R\indices{_a_b^e^f}R\indices{_c_e^g^h}R_{dfgh}
        + e_{24} R^{abcd}R\indices{_a_b^e^f}R\indices{_c^g_e^h}R_{dgfh}
        \nonumber\\& \quad + e_{25} R^{abcd}R\indices{_a^e_c^f}R\indices{_e^g_f^h}R_{bgdh}
        + e_{26} R^{abcd}R\indices{_a^e_c^f}R\indices{_e^g_b^h}R_{fgdh}
\end{align}
There are totally 12 dimension-generic solutions, their coefficients $e_{i, j}$ are listed in table \ref{tab:quartic-qtg-dimgen-sol} below, where $i$ labels different solutions and $j$ labels the 26 coefficients of one solution.
\begin{longtable}{|c|l|}
    \hline
    $e_{1, 1}$ & $\frac{2 D^9-61 D^8+773 D^7-5451 D^6+23821 D^5-67174 D^4+121930 D^3-135736 D^2+81920 D-19456}{2 (D-4) (D-3) (D-2)^5 (D-1) D \left(D^3-9 D^2+26 D-22\right)}$ \\
    $e_{1, 2}$ & $\frac{-3 D^{10}+94 D^9-1237 D^8+9168 D^7-42780 D^6+131846 D^5-271580 D^4+367060 D^3-308704 D^2+145024 D-29184}{(D-4) (D-3) (D-2)^5 (D-1) D \left(D^3-9 D^2+26 D-22\right)}$ \\
    $e_{1, 3}$ & $\frac{4 \left(2 D^9-59 D^8+733 D^7-5103 D^6+22103 D^5-61918 D^4+111738 D^3-123512 D^2+73632 D-17024\right)}{(D-4) (D-3) (D-2)^4 (D-1) D \left(D^3-9 D^2+26 D-22\right)}$ \\
    $e_{1, 4}$ & $\frac{2 D^8-56 D^7+633 D^6-3948 D^5+15253 D^4-37812 D^3+58752 D^2-51840 D+19456}{(D-4) (D-3) (D-2)^5 (D-1) D}$ \\
    $e_{1, 5}$ & $\frac{-2 D^7+45 D^6-398 D^5+1895 D^4-5476 D^3+9776 D^2-10112 D+4864}{(D-4) (D-3) (D-2)^4 (D-1) D}$ \\
    $e_{1, 6}$ & $\frac{4 \left(2 D^8-59 D^7+730 D^6-5041 D^5+21496 D^4-58348 D^3+98636 D^2-94560 D+38912\right)}{(D-4) (D-3) (D-2)^3 (D-1) D \left(D^3-9 D^2+26 D-22\right)}$ \\
    $e_{1, 7}$ & $-\frac{4 \left(D^6-19 D^5+131 D^4-409 D^3+520 D^2+88 D-608\right)}{(D-4) (D-3) (D-2)^3 (D-1) D}$ \\
    $e_{1, 8}$ & $\frac{D^{10}-36 D^9+557 D^8-4979 D^7+28834 D^6-113919 D^5+312276 D^4-587102 D^3+723424 D^2-525760 D+170240}{2 (D-4) (D-3) (D-2)^4 (D-1) D \left(D^3-9 D^2+26 D-22\right)}$ \\
    $e_{1, 9}$ & $-\frac{2 \left(D^8-30 D^7+376 D^6-2636 D^5+11493 D^4-32254 D^3+57146 D^2-58160 D+25536\right)}{(D-4) (D-3) (D-2)^2 (D-1) D \left(D^3-9 D^2+26 D-22\right)}$ \\
    $e_{1, 10}$ & $-\frac{D^8-30 D^7+353 D^6-2250 D^5+8748 D^4-21514 D^3+32672 D^2-27712 D+9728}{2 (D-4) (D-3) (D-2)^4 (D-1) D}$ \\
    $e_{1, 11}$ & $\frac{D^6-23 D^5+199 D^4-861 D^3+1996 D^2-2216 D+608}{(D-4) (D-3) (D-2)^2 (D-1) D}$ \\
    $e_{1, 15}$ & $\frac{D^3-12 D^2+41 D-38}{(D-1) D \left(D^3-9 D^2+26 D-22\right)}$ \\
    $e_{1, 17}$ & $\frac{76-20 D}{D^3-5 D^2+6 D}$ \\
    $e_{1, 20}$ & $-\frac{D^5-10 D^4+28 D^3+18 D^2-173 D+160}{8 (D-3) (D-2)^3 (D-1)}$ \\
    $e_{1, 26}$ & $1$ \\
    \hline
    $e_{2, 1}$ & $\frac{12 D^8-244 D^7+2138 D^6-10521 D^5+31695 D^4-59494 D^3+67138 D^2-40696 D+9728}{(D-4) (D-3) (D-2)^5 (D-1) D \left(D^3-9 D^2+26 D-22\right)}$ \\
    $e_{2, 2}$ & $-\frac{4 \left(9 D^9-191 D^8+1766 D^7-9316 D^6+30809 D^5-65961 D^4+90838 D^3-76978 D^2+36256 D-7296\right)}{(D-4) (D-3) (D-2)^5 (D-1) D \left(D^3-9 D^2+26 D-22\right)}$ \\
    $e_{2, 3}$ & $\frac{8 \left(11 D^8-224 D^7+1964 D^6-9662 D^5+29067 D^4-54398 D^3+61026 D^2-36552 D+8512\right)}{(D-4) (D-3) (D-2)^4 (D-1) D \left(D^3-9 D^2+26 D-22\right)}$ \\
    $e_{2, 4}$ & $\frac{2 \left(13 D^7-228 D^6+1730 D^5-7392 D^4+19201 D^3-30196 D^2+26400 D-9728\right)}{(D-4) (D-3) (D-2)^5 (D-1) D}$ \\
    $e_{2, 5}$ & $-\frac{2 \left(14 D^6-185 D^5+1018 D^4-3059 D^3+5380 D^2-5344 D+2432\right)}{(D-4) (D-3) (D-2)^4 (D-1) D}$ \\
    $e_{2, 6}$ & $-\frac{4 \left(D^8-45 D^7+671 D^6-5095 D^5+22712 D^4-62348 D^3+104224 D^2-97464 D+38912\right)}{(D-4) (D-3) (D-2)^3 (D-1) D \left(D^3-9 D^2+26 D-22\right)}$ \\
    $e_{2, 7}$ & $-\frac{8 \left(4 D^5-41 D^4+144 D^3-167 D^2-116 D+304\right)}{(D-4) (D-3) (D-2)^3 (D-1) D}$ \\
    $e_{2, 8}$ & $\frac{7 D^9-178 D^8+2013 D^7-13299 D^6+56610 D^5-161107 D^4+306544 D^3-375726 D^2+268688 D-85120}{(D-4) (D-3) (D-2)^4 (D-1) D \left(D^3-9 D^2+26 D-22\right)}$ \\
    $e_{2, 9}$ & $-\frac{8 \left(3 D^7-62 D^6+551 D^5-2733 D^4+8181 D^3-14787 D^2+14903 D-6384\right)}{(D-4) (D-3) (D-2)^2 (D-1) D \left(D^3-9 D^2+26 D-22\right)}$ \\
    $e_{2, 10}$ & $-\frac{2 \left(3 D^7-56 D^6+445 D^5-1959 D^4+5160 D^3-8093 D^2+6928 D-2432\right)}{(D-4) (D-3) (D-2)^4 (D-1) D}$ \\
    $e_{2, 11}$ & $\frac{2 \left(7 D^5-89 D^4+437 D^3-1031 D^2+1108 D-304\right)}{(D-4) (D-3) (D-2)^2 (D-1) D}$ \\
    $e_{2, 15}$ & $\frac{-D^3+12 D^2-41 D+38}{D^5-10 D^4+35 D^3-48 D^2+22 D}$ \\
    $e_{2, 17}$ & $-\frac{4 \left(D^2-8 D+19\right)}{(D-3) (D-2) D}$ \\
    $e_{2, 20}$ & $-\frac{D^5-12 D^4+61 D^3-167 D^2+242 D-137}{4 (D-3) (D-2)^3 (D-1)}$ \\
    $e_{2, 25}$ & $1$ \\
    \hline
    $e_{3, 1}$ & $\frac{D^9-30 D^8+380 D^7-2695 D^6+11858 D^5-33610 D^4+61132 D^3-68000 D^2+40960 D-9728}{(D-4) (D-3) (D-2)^5 (D-1) D \left(D^3-9 D^2+26 D-22\right)}$ \\
    $e_{3, 2}$ & $\frac{-3 D^{10}+92 D^9-1207 D^8+8988 D^7-42238 D^6+131008 D^5-270984 D^4+366928 D^3-308704 D^2+145024 D-29184}{(D-4) (D-3) (D-2)^5 (D-1) D \left(D^3-9 D^2+26 D-22\right)}$ \\
    $e_{3, 3}$ & $\frac{8 \left(D^9-29 D^8+360 D^7-2521 D^6+10999 D^5-30982 D^4+56036 D^3-61888 D^2+36816 D-8512\right)}{(D-4) (D-3) (D-2)^4 (D-1) D \left(D^3-9 D^2+26 D-22\right)}$ \\
    $e_{3, 4}$ & $\frac{2 \left(D^8-27 D^7+307 D^6-1947 D^5+7638 D^4-19104 D^3+29728 D^2-26112 D+9728\right)}{(D-4) (D-3) (D-2)^5 (D-1) D}$ \\
    $e_{3, 5}$ & $-\frac{2 \left(D^7-22 D^6+199 D^5-970 D^4+2832 D^3-5040 D^2+5152 D-2432\right)}{(D-4) (D-3) (D-2)^4 (D-1) D}$ \\
    $e_{3, 6}$ & $\frac{4 \left(2 D^8-59 D^7+736 D^6-5129 D^5+22006 D^4-59796 D^3+100632 D^2-95616 D+38912\right)}{(D-4) (D-3) (D-2)^3 (D-1) D \left(D^3-9 D^2+26 D-22\right)}$ \\
    $e_{3, 7}$ & $-\frac{4 \left(D^6-18 D^5+123 D^4-382 D^3+468 D^2+136 D-608\right)}{(D-4) (D-3) (D-2)^3 (D-1) D}$ \\
    $e_{3, 8}$ & $\frac{D^{10}-35 D^9+541 D^8-4887 D^7+28686 D^6-114666 D^5+316648 D^4-596688 D^3+733520 D^2-529984 D+170240}{2 (D-4) (D-3) (D-2)^4 (D-1) D \left(D^3-9 D^2+26 D-22\right)}$ \\
    $e_{3, 9}$ & $-\frac{2 \left(D^2-7 D+14\right) \left(D^6-22 D^5+195 D^4-906 D^3+2364 D^2-3280 D+1824\right)}{(D-4) (D-3) (D-2)^2 (D-1) D \left(D^3-9 D^2+26 D-22\right)}$ \\
    $e_{3, 10}$ & $-\frac{D^8-28 D^7+329 D^6-2138 D^5+8496 D^4-21248 D^3+32576 D^2-27712 D+9728}{2 (D-4) (D-3) (D-2)^4 (D-1) D}$ \\
    $e_{3, 11}$ & $\frac{D^6-22 D^5+195 D^4-866 D^3+2020 D^2-2216 D+608}{(D-4) (D-3) (D-2)^2 (D-1) D}$ \\
    $e_{3, 15}$ & $\frac{D^3-12 D^2+41 D-38}{(D-1) D \left(D^3-9 D^2+26 D-22\right)}$ \\
    $e_{3, 17}$ & $\frac{2 \left(D^2-13 D+38\right)}{D \left(D^2-5 D+6\right)}$ \\
    $e_{3, 20}$ & $-\frac{(D-4) \left(D^3-10 D^2+31 D-26\right)}{4 (D-3) (D-2)^3 (D-1)}$ \\
    $e_{3, 24}$ & $1$ \\
    \hline
    $e_{4, 1}$ & $\frac{2 \left(11 D^5-140 D^4+668 D^3-1502 D^2+1544 D-512\right)}{(D-4) (D-3) (D-2)^5 (D-1) D}$ \\
    $e_{4, 2}$ & $-\frac{16 \left(4 D^6-55 D^5+293 D^4-780 D^3+1083 D^2-728 D+192\right)}{(D-4) (D-3) (D-2)^5 (D-1) D}$ \\
    $e_{4, 3}$ & $\frac{32 \left(5 D^5-64 D^4+306 D^3-687 D^2+700 D-224\right)}{(D-4) (D-3) (D-2)^4 (D-1) D}$ \\
    $e_{4, 4}$ & $\frac{8 \left(5 D^7-100 D^6+838 D^5-3856 D^4+10553 D^3-17148 D^2+15232 D-5632\right)}{(D-4) (D-3) (D-2)^5 (D-1) D}$ \\
    $e_{4, 5}$ & $-\frac{8 \left(6 D^6-89 D^5+530 D^4-1675 D^3+3036 D^2-3072 D+1408\right)}{(D-4) (D-3) (D-2)^4 (D-1) D}$ \\
    $e_{4, 6}$ & $\frac{64 \left(3 D^4-37 D^3+167 D^2-335 D+256\right)}{(D-4) (D-3) (D-2)^3 (D-1) D}$ \\
    $e_{4, 7}$ & $-\frac{32 \left(2 D^5-23 D^4+88 D^3-111 D^2-60 D+176\right)}{(D-4) (D-3) (D-2)^3 (D-1) D}$ \\
    $e_{4, 8}$ & $\frac{2 \left(5 D^6-99 D^5+778 D^4-3182 D^3+7250 D^2-8800 D+4480\right)}{(D-4) (D-3) (D-2)^4 (D-1) D}$ \\
    $e_{4, 9}$ & $-\frac{16 (2 D-7) \left(D^3-12 D^2+41 D-48\right)}{(D-4) (D-3) (D-2)^2 (D-1) D}$ \\
    $e_{4, 10}$ & $-\frac{8 \left(D^6-22 D^5+188 D^4-827 D^3+2013 D^2-2600 D+1408\right)}{(D-4) (D-3) (D-2)^4 D}$ \\
    $e_{4, 11}$ & $\frac{8 \left(3 D^5-45 D^4+245 D^3-607 D^2+652 D-176\right)}{(D-4) (D-3) (D-2)^2 (D-1) D}$ \\
    $e_{4, 15}$ & $\frac{4}{D-D^2}$ \\
    $e_{4, 17}$ & $-\frac{16 \left(D^2-6 D+11\right)}{(D-3) (D-2) D}$ \\
    $e_{4, 20}$ & $-\frac{D^5-14 D^4+79 D^3-224 D^2+316 D-170}{2 (D-3) (D-2)^3 (D-1)}$ \\
    $e_{4, 23}$ & $1$ \\
    \hline
    $e_{5, 1}$ & $\frac{4 \left(11 D^5-140 D^4+668 D^3-1502 D^2+1544 D-512\right)}{(D-4) (D-3) (D-2)^5 (D-1) D}$ \\
    $e_{5, 2}$ & $-\frac{32 \left(4 D^6-55 D^5+293 D^4-780 D^3+1083 D^2-728 D+192\right)}{(D-4) (D-3) (D-2)^5 (D-1) D}$ \\
    $e_{5, 3}$ & $\frac{64 \left(5 D^5-64 D^4+306 D^3-687 D^2+700 D-224\right)}{(D-4) (D-3) (D-2)^4 (D-1) D}$ \\
    $e_{5, 4}$ & $\frac{16 \left(5 D^7-100 D^6+838 D^5-3856 D^4+10553 D^3-17148 D^2+15232 D-5632\right)}{(D-4) (D-3) (D-2)^5 (D-1) D}$ \\
    $e_{5, 5}$ & $-\frac{16 \left(6 D^6-89 D^5+530 D^4-1675 D^3+3036 D^2-3072 D+1408\right)}{(D-4) (D-3) (D-2)^4 (D-1) D}$ \\
    $e_{5, 6}$ & $\frac{128 \left(3 D^4-37 D^3+167 D^2-335 D+256\right)}{(D-4) (D-3) (D-2)^3 (D-1) D}$ \\
    $e_{5, 7}$ & $-\frac{64 \left(2 D^5-23 D^4+88 D^3-111 D^2-60 D+176\right)}{(D-4) (D-3) (D-2)^3 (D-1) D}$ \\
    $e_{5, 8}$ & $\frac{4 \left(5 D^6-99 D^5+778 D^4-3182 D^3+7250 D^2-8800 D+4480\right)}{(D-4) (D-3) (D-2)^4 (D-1) D}$ \\
    $e_{5, 9}$ & $-\frac{32 (2 D-7) \left(D^3-12 D^2+41 D-48\right)}{(D-4) (D-3) (D-2)^2 (D-1) D}$ \\
    $e_{5, 10}$ & $-\frac{16 \left(D^6-22 D^5+188 D^4-827 D^3+2013 D^2-2600 D+1408\right)}{(D-4) (D-3) (D-2)^4 D}$ \\
    $e_{5, 11}$ & $\frac{16 \left(3 D^5-45 D^4+245 D^3-607 D^2+652 D-176\right)}{(D-4) (D-3) (D-2)^2 (D-1) D}$ \\
    $e_{5, 15}$ & $\frac{8}{D-D^2}$ \\
    $e_{5, 17}$ & $-\frac{32 \left(D^2-6 D+11\right)}{(D-3) (D-2) D}$ \\
    $e_{5, 20}$ & $\frac{-D^5+14 D^4-79 D^3+224 D^2-316 D+170}{(D-3) (D-2)^3 (D-1)}$ \\
    $e_{5, 22}$ & $1$ \\
    \hline
    $e_{6, 1}$ & $-\frac{2 D \left(D^2-D-4\right)}{(D-4) (D-2)^5 (D-1)}$ \\
    $e_{6, 2}$ & $\frac{8 D^2 \left(D^2-3 D+1\right)}{(D-4) (D-2)^5 (D-1)}$ \\
    $e_{6, 3}$ & $-\frac{16 D \left(D^2-D-4\right)}{(D-4) (D-2)^4 (D-1)}$ \\
    $e_{6, 4}$ & $-\frac{4 \left(2 D^5-13 D^4+15 D^3+68 D^2-192 D+128\right)}{(D-4) (D-2)^5 (D-1)}$ \\
    $e_{6, 5}$ & $\frac{4 \left(D^4+3 D^3-36 D^2+80 D-64\right)}{(D-4) (D-2)^4 (D-1)}$ \\
    $e_{6, 6}$ & $-\frac{32 (3 D-8)}{(D-4) (D-2)^3 (D-1)}$ \\
    $e_{6, 7}$ & $\frac{16 \left(D^3-5 D^2+12 D-16\right)}{(D-4) (D-2)^3 (D-1)}$ \\
    $e_{6, 8}$ & $-\frac{2 \left(D^4-4 D^3-9 D^2+56 D-64\right)}{(D-4) (D-2)^4 (D-1)}$ \\
    $e_{6, 9}$ & $\frac{8 \left(D^2-D-8\right)}{(D-4) (D-2)^2 (D-1)}$ \\
    $e_{6, 10}$ & $\frac{4 D \left(D^4-9 D^3+29 D^2-39 D+16\right)}{(D-4) (D-2)^4 (D-1)}$ \\
    $e_{6, 11}$ & $-\frac{4 D \left(D^2-D-8\right)}{(D-4) (D-2)^2 (D-1)}$ \\
    $e_{6, 17}$ & $-\frac{8}{D-2}$ \\
    $e_{6, 20}$ & $-\frac{D^3-8 D^2+21 D-16}{2 (D-2)^3}$ \\
    $e_{6, 21}$ & $1$ \\
    \hline
    $e_{7, 1}$ & $-\frac{D^6-16 D^5+103 D^4-336 D^3+570 D^2-458 D+128}{(D-4) (D-3) (D-2)^2 (D-1) D \left(D^3-9 D^2+26 D-22\right)}$ \\
    $e_{7, 2}$ & $\frac{3 D^7-51 D^6+356 D^5-1305 D^4+2665 D^3-2984 D^2+1692 D-384}{(D-4) (D-3) (D-2)^2 (D-1) D \left(D^3-9 D^2+26 D-22\right)}$ \\
    $e_{7, 3}$ & $\frac{-7 D^6+114 D^5-743 D^4+2440 D^3-4140 D^2+3296 D-896}{(D-4) (D-3) (D-2) (D-1) D \left(D^3-9 D^2+26 D-22\right)}$ \\
    $e_{7, 4}$ & $-\frac{2 \left(D^2-5 D+8\right) \left(D^3-9 D^2+23 D-16\right)}{(D-4) (D-3) (D-2)^2 (D-1) D}$ \\
    $e_{7, 5}$ & $\frac{2 \left(D^4-10 D^3+33 D^2-48 D+32\right)}{D \left(D^4-10 D^3+35 D^2-50 D+24\right)}$ \\
    $e_{7, 6}$ & $\frac{-9 D^5+139 D^4-851 D^3+2561 D^2-3728 D+2048}{(D-4) (D-3) (D-1) D \left(D^3-9 D^2+26 D-22\right)}$ \\
    $e_{7, 7}$ & $\frac{2 \left(D^3-7 D^2+6 D+16\right)}{(D-4) (D-3) (D-1) D}$ \\
    $e_{7, 8}$ & $-\frac{D^7-22 D^6+206 D^5-1059 D^4+3213 D^3-5733 D^2+5554 D-2240}{2 (D-4) (D-3) (D-2) (D-1) D \left(D^3-9 D^2+26 D-22\right)}$ \\
    $e_{7, 9}$ & $\frac{(D-2) \left(4 D^5-65 D^4+428 D^3-1409 D^2+2258 D-1344\right)}{2 (D-4) (D-3) (D-1) D \left(D^3-9 D^2+26 D-22\right)}$ \\
    $e_{7, 10}$ & $\frac{D^5-15 D^4+85 D^3-229 D^2+290 D-128}{2 (D-4) (D-3) (D-2) (D-1) D}$ \\
    $e_{7, 11}$ & $-\frac{(D-2) \left(D^3-10 D^2+25 D-8\right)}{(D-4) (D-3) (D-1) D}$ \\
    $e_{7, 15}$ & $\frac{3 D^2-15 D+16}{-4 D^4+36 D^3-104 D^2+88 D}$ \\
    $e_{7, 17}$ & $\frac{4-D}{(D-3) D}$ \\
    $e_{7, 19}$ & $1$ \\
    \hline
    $e_{8, 1}$ & $\frac{2 \left(3 D^2-9 D+4\right)}{(D-3) (D-2)^2 (D-1) D}$ \\
    $e_{8, 2}$ & $\frac{-18 D^3+70 D^2-72 D+24}{(D-3) (D-2)^2 (D-1) D}$ \\
    $e_{8, 3}$ & $\frac{4 \left(11 D^2-33 D+14\right)}{(D-3) (D-2) (D-1) D}$ \\
    $e_{8, 4}$ & $\frac{4 \left(3 D^4-28 D^3+101 D^2-160 D+88\right)}{(D-3) (D-2)^2 (D-1) D}$ \\
    $e_{8, 5}$ & $-\frac{4 \left(3 D^3-14 D^2+25 D-22\right)}{(D-3) (D-2) (D-1) D}$ \\
    $e_{8, 6}$ & $\frac{16 (3 D-8)}{(D-3) (D-1) D}$ \\
    $e_{8, 7}$ & $\frac{-20 D^2+40 D+44}{D^3-4 D^2+3 D}$ \\
    $e_{8, 8}$ & $\frac{3 D^3-26 D^2+73 D-70}{D \left(D^3-6 D^2+11 D-6\right)}$ \\
    $e_{8, 9}$ & $-\frac{2 (D-2) (5 D-21)}{(D-3) (D-1) D}$ \\
    $e_{8, 10}$ & $\frac{-3 D^4+32 D^3-119 D^2+182 D-88}{D \left(D^3-6 D^2+11 D-6\right)}$ \\
    $e_{8, 11}$ & $\frac{6 D^3-42 D^2+74 D-22}{D \left(D^2-4 D+3\right)}$ \\
    $e_{8, 15}$ & $-\frac{1}{D}$ \\
    $e_{8, 17}$ & $-\frac{2 \left(D^2-6 D+11\right)}{(D-3) D}$ \\
    $e_{8, 18}$ & $1$ \\
    \hline
    $e_{9, 1}$ & $\frac{1}{-D^3+9 D^2-26 D+22}$ \\
    $e_{9, 2}$ & $\frac{3 (D-1)}{D^3-9 D^2+26 D-22}$ \\
    $e_{9, 3}$ & $\frac{14-6 D}{D^3-9 D^2+26 D-22}$ \\
    $e_{9, 6}$ & $-\frac{3 \left(D^2-5 D+8\right)}{D^3-9 D^2+26 D-22}$ \\
    $e_{9, 8}$ & $-\frac{3 (D-3)}{2 \left(D^3-9 D^2+26 D-22\right)}$ \\
    $e_{9, 9}$ & $\frac{3 (D-3) (D-1)}{D^3-9 D^2+26 D-22}$ \\
    $e_{9, 15}$ & $\frac{3 D^2-15 D+16}{-4 D^3+36 D^2-104 D+88}$ \\
    $e_{9, 16}$ & $1$ \\
    \hline
    $e_{10, 1}$ & $\frac{D^2-4 D+2}{2 (D-4) (D-2)^2 (D-1)}$ \\
    $e_{10, 2}$ & $-\frac{(3 D-2) \left(D^2-4 D+2\right)}{2 (D-4) (D-2)^2 (D-1)}$ \\
    $e_{10, 3}$ & $\frac{4 \left(D^2-4 D+2\right)}{(D-4) (D-2) (D-1)}$ \\
    $e_{10, 4}$ & $\frac{D^4-10 D^3+39 D^2-68 D+40}{(D-4) (D-2)^2 (D-1)}$ \\
    $e_{10, 5}$ & $\frac{-D^3+5 D^2-8 D+8}{(D-4) (D-2) (D-1)}$ \\
    $e_{10, 6}$ & $\frac{2 (2 D-7)}{D^2-5 D+4}$ \\
    $e_{10, 7}$ & $\frac{-2 D^2+6 D+4}{D^2-5 D+4}$ \\
    $e_{10, 8}$ & $\frac{(D-3) \left(D^2-6 D+10\right)}{4 (D-4) (D-2) (D-1)}$ \\
    $e_{10, 9}$ & $-\frac{(D-3)^2}{(D-4) (D-1)}$ \\
    $e_{10, 10}$ & $-\frac{(D-3) \left(D^3-8 D^2+20 D-12\right)}{4 (D-4) (D-2) (D-1)}$ \\
    $e_{10, 11}$ & $\frac{D^3-8 D^2+19 D-8}{2 (D-4) (D-1)}$ \\
    $e_{10, 14}$ & $1$ \\
    \hline
    $e_{11, 1}$ & $\frac{D}{2 (D-4) (D-2)^2 (D-1)}$ \\
    $e_{11, 2}$ & $-\frac{D (3 D-2)}{2 (D-4) (D-2)^2 (D-1)}$ \\
    $e_{11, 3}$ & $\frac{4 D}{(D-4) (D-2) (D-1)}$ \\
    $e_{11, 4}$ & $\frac{2 \left(D^3-7 D^2+17 D-12\right)}{(D-4) (D-2)^2 (D-1)}$ \\
    $e_{11, 5}$ & $-\frac{2 \left(D^2-3 D+4\right)}{D^3-7 D^2+14 D-8}$ \\
    $e_{11, 6}$ & $-\frac{2 (D-6)}{D^2-5 D+4}$ \\
    $e_{11, 7}$ & $-\frac{8}{D^2-5 D+4}$ \\
    $e_{11, 8}$ & $\frac{(D-3) (3 D-8)}{4 (D-4) (D-2) (D-1)}$ \\
    $e_{11, 9}$ & $\frac{6-2 D}{D^2-5 D+4}$ \\
    $e_{11, 10}$ & $-\frac{(D-3) \left(3 D^2-12 D+8\right)}{4 (D-4) (D-2) (D-1)}$ \\
    $e_{11, 11}$ & $\frac{(D-3) D}{(D-4) (D-1)}$ \\
    $e_{11, 13}$ & $1$ \\
    \hline
    $e_{12, 1}$ & $\frac{D^2-4 D+2}{(D-4) (D-2)^2 (D-1)}$ \\
    $e_{12, 2}$ & $-\frac{(3 D-2) \left(D^2-4 D+2\right)}{(D-4) (D-2)^2 (D-1)}$ \\
    $e_{12, 3}$ & $\frac{8 \left(D^2-4 D+2\right)}{(D-4) (D-2) (D-1)}$ \\
    $e_{12, 4}$ & $\frac{2 \left(D^4-10 D^3+39 D^2-68 D+40\right)}{(D-4) (D-2)^2 (D-1)}$ \\
    $e_{12, 5}$ & $-\frac{2 \left(D^3-5 D^2+8 D-8\right)}{(D-4) (D-2) (D-1)}$ \\
    $e_{12, 6}$ & $\frac{4 (2 D-7)}{D^2-5 D+4}$ \\
    $e_{12, 7}$ & $\frac{-4 D^2+12 D+8}{D^2-5 D+4}$ \\
    $e_{12, 8}$ & $\frac{(D-3) \left(D^2-6 D+10\right)}{2 (D-4) (D-2) (D-1)}$ \\
    $e_{12, 9}$ & $-\frac{2 (D-3)^2}{(D-4) (D-1)}$ \\
    $e_{12, 10}$ & $-\frac{(D-3) \left(D^3-8 D^2+20 D-12\right)}{2 (D-4) (D-2) (D-1)}$ \\
    $e_{12, 11}$ & $\frac{D^3-7 D^2+14 D-4}{(D-4) (D-1)}$ \\
    $e_{12, 12}$ & $1$ \\
    \hline
    \caption{Coefficients of dimension-generic quartic \ac{sqt} gravity solutions. Zero coefficients are omitted.}
    \label{tab:quartic-qtg-dimgen-sol}
\end{longtable}
There are also dimension-specific solutions for $D = 3$ and $D = 4$, given in table \ref{tab:quartic-qtg-sol-3d} and \ref{tab:quartic-qtg-sol-4d} respectively.
\begin{longtable}{|l|}
    \hline
    $\{e_i\}$ \\
    $(\frac{1}{6},-\frac{1}{2},\frac{4}{3},-\frac{3}{2},0,0,0,0,0,0,0,0,0,0,0,0,0,0,0,0,0,0,0,0,0,1)$ \\
    $(\frac{1}{6},0,\frac{4}{3},-3,0,0,0,0,0,0,0,0,0,0,0,0,0,0,0,0,0,0,0,0,1,0)$ \\
    $(\frac{1}{12},-\frac{1}{2},\frac{2}{3},0,0,0,0,0,0,0,0,0,0,0,0,0,0,0,0,0,0,0,0,1,0,0)$ \\
    $(\frac{7}{6},-4,\frac{16}{3},-4,0,0,0,0,0,0,0,0,0,0,0,0,0,0,0,0,0,0,1,0,0,0)$ \\
    $(\frac{7}{3},-8,\frac{32}{3},-8,0,0,0,0,0,0,0,0,0,0,0,0,0,0,0,0,0,1,0,0,0,0)$ \\
    $(\frac{1}{3},0,\frac{8}{3},-6,0,0,0,0,0,0,0,0,0,0,0,0,0,0,0,0,1,0,0,0,0,0)$ \\
    $(-1,8,0,-16,0,0,0,0,0,0,0,0,0,0,0,0,0,0,0,1,0,0,0,0,0,0)$ \\
    $(\frac{1}{12},-\frac{1}{2},\frac{2}{3},0,0,0,0,0,0,0,0,0,0,0,0,0,0,0,1,0,0,0,0,0,0,0)$ \\
    $(\frac{4}{3},-5,\frac{14}{3},-2,0,0,0,0,0,0,0,0,0,0,0,0,0,1,0,0,0,0,0,0,0,0)$ \\
    $(\frac{1}{3},-1,\frac{5}{3},-2,0,0,0,0,0,0,0,0,0,0,0,0,1,0,0,0,0,0,0,0,0,0)$ \\
    $(\frac{1}{4},-\frac{3}{2},2,0,0,0,0,0,0,0,0,0,0,0,0,1,0,0,0,0,0,0,0,0,0,0)$ \\
    $(3,-12,8,0,0,0,0,0,0,0,0,0,0,0,1,0,0,0,0,0,0,0,0,0,0,0)$ \\
    $(\frac{7}{12},-\frac{9}{4},\frac{5}{3},-\frac{1}{2},0,0,0,0,0,0,0,0,0,1,0,0,0,0,0,0,0,0,0,0,0,0)$ \\
    $(\frac{1}{12},-\frac{1}{4},\frac{2}{3},-1,0,0,0,0,0,0,0,0,1,0,0,0,0,0,0,0,0,0,0,0,0,0)$ \\
    $(\frac{5}{6},-\frac{7}{2},\frac{8}{3},0,0,0,0,0,0,0,0,1,0,0,0,0,0,0,0,0,0,0,0,0,0,0)$ \\
    $(\frac{1}{3},-1,\frac{2}{3},-1,0,0,0,0,0,0,1,0,0,0,0,0,0,0,0,0,0,0,0,0,0,0)$ \\
    $(0,1,0,-4,0,0,0,0,0,1,0,0,0,0,0,0,0,0,0,0,0,0,0,0,0,0)$ \\
    $(1,-4,2,0,0,0,0,0,1,0,0,0,0,0,0,0,0,0,0,0,0,0,0,0,0,0)$ \\
    $(1,-4,0,0,0,0,0,1,0,0,0,0,0,0,0,0,0,0,0,0,0,0,0,0,0,0)$ \\
    $(\frac{1}{3},-\frac{3}{2},\frac{7}{6},0,0,0,1,0,0,0,0,0,0,0,0,0,0,0,0,0,0,0,0,0,0,0)$ \\
    $(\frac{1}{2},-\frac{5}{2},2,0,0,1,0,0,0,0,0,0,0,0,0,0,0,0,0,0,0,0,0,0,0,0)$ \\
    $(-\frac{1}{6},1,-\frac{4}{3},-\frac{1}{2},1,0,0,0,0,0,0,0,0,0,0,0,0,0,0,0,0,0,0,0,0,0)$ \\
    \hline
    \caption{Coefficients of quartic \ac{sqt} gravity for $D = 3$}
    \label{tab:quartic-qtg-sol-3d}
\end{longtable}

\begin{longtable}{|l|}
    \hline
    $\{e_i\}$ \\
    $(\frac{35}{48},-\frac{23}{4},\frac{20}{3},\frac{5}{2},-4,3,0,\frac{1}{4},0,-\frac{1}{4},0,-2,0,0,-\frac{1}{12},0,0,0,0,-\frac{1}{16},0,0,0,0,0,1)$ \\
    $(\frac{19}{32},-5,\frac{20}{3},\frac{13}{4},-\frac{11}{2},2,0,\frac{3}{16},0,-\frac{1}{4},0,-\frac{5}{2},0,0,\frac{1}{12},0,0,0,0,-\frac{5}{32},0,0,0,0,1,0)$ \\
    $(\frac{11}{24},-\frac{7}{2},4,1,-2,2,0,\frac{1}{8},0,0,0,-1,0,0,-\frac{1}{12},0,0,0,0,0,0,0,0,1,0,0)$ \\
    $(\frac{41}{24},-14,16,7,-10,8,0,\frac{3}{4},0,-1,0,-6,0,0,-\frac{1}{3},0,0,0,0,-\frac{1}{8},0,0,1,0,0,0)$ \\
    $(\frac{41}{12},-28,32,14,-20,16,0,\frac{3}{2},0,-2,0,-12,0,0,-\frac{2}{3},0,0,0,0,-\frac{1}{4},0,1,0,0,0,0)$ \\
    $(\frac{13}{12},-9,\frac{32}{3},6,-8,4,0,\frac{1}{2},0,-1,0,-4,0,0,0,0,0,0,0,-\frac{1}{4},1,0,0,0,0,0)$ \\
    $(\frac{1}{3},-\frac{5}{2},\frac{8}{3},\frac{1}{2},-1,\frac{3}{2},0,\frac{1}{8},0,0,0,-\frac{1}{2},0,0,-\frac{1}{8},0,0,0,1,0,0,0,0,0,0,0)$ \\
    $(\frac{5}{8},-\frac{21}{4},6,3,-4,3,0,\frac{3}{8},0,-\frac{3}{4},0,-3,0,0,-\frac{1}{4},0,0,1,0,0,0,0,0,0,0,0)$ \\
    $(\frac{7}{12},-\frac{19}{4},\frac{17}{3},3,-4,2,0,\frac{1}{4},0,-\frac{3}{4},0,-2,0,0,0,0,1,0,0,0,0,0,0,0,0,0)$ \\
    $(\frac{5}{8},-\frac{9}{2},4,0,0,3,0,\frac{3}{8},0,0,0,0,0,0,-\frac{1}{2},1,0,0,0,0,0,0,0,0,0,0)$ \\
    $(-\frac{1}{12},\frac{5}{8},-\frac{2}{3},0,0,-\frac{1}{2},0,0,0,-\frac{1}{8},0,-\frac{1}{2},0,1,0,0,0,0,0,0,0,0,0,0,0,0)$ \\
    $(\frac{3}{8},-3,4,\frac{3}{2},-3,1,0,\frac{1}{8},0,-\frac{1}{4},0,-1,1,0,0,0,0,0,0,0,0,0,0,0,0,0)$ \\
    $(-\frac{1}{6},\frac{5}{4},-\frac{4}{3},0,0,-1,0,0,0,-\frac{1}{4},1,0,0,0,0,0,0,0,0,0,0,0,0,0,0,0)$ \\
    $(-\frac{1}{4},2,-2,0,0,-2,0,-\frac{1}{4},1,0,0,0,0,0,0,0,0,0,0,0,0,0,0,0,0,0)$ \\
    $(-\frac{1}{12},\frac{3}{4},-\frac{7}{6},-\frac{1}{2},1,-\frac{1}{2},1,0,0,0,0,0,0,0,0,0,0,0,0,0,0,0,0,0,0,0)$ \\
    \hline
    \caption{Coefficients of quartic \ac{sqt} gravity for $D = 4$}
    \label{tab:quartic-qtg-sol-4d}
\end{longtable}
However, the $D = 3$ solutions must be trivial otherwise it contradicts with the observation made in \cite{aspects-of-3d-higher-grav}. To check this, we can expand the solutions into the six independent components of the three dimensional Riemann tensor under some tetrad
\begin{equation}\label{eq:3d-riem-comps}
    R_{1212}, \, R_{1213}, \, R_{1223}, \, R_{1313}, \, R_{1323}, \, R_{2323}
\end{equation}
Although on a generic tetrad this introduces metric components, which are not independent from the Riemann tensor, the metric components can be treated as the flat one $\delta_{ij}$. This is because it's always possible to choose an orthonormal tetrad where the metric component matrix is diagonalized with components being either $1$ or $-1$
\begin{equation}
    \dd s^2 = g_{ij} \dd x^i \dd x^j = \eta_{ij} \theta^i \theta^j
\end{equation}
where $\theta^i$ is the dual tetrad, $\eta = \mathrm{diag}(-1, \cdots, 1, \cdots)$. We can then further transform $\eta_{ij}$ into $\delta_{ij}$ by multiplying some basis vectors by a factor of $i$, similar to Wick rotation. This turns some components in \eqref{eq:3d-riem-comps} into imaginary numbers, but does not affect our results. Applying this method to the solution in table \ref{tab:quartic-qtg-sol-3d} and also the quintic case below, we found the $D = 3$ solutions are all trivial, that is, vanish identically when evaluated on any $D = 3$ metric.

\subsection{Quintic order}
For the quintic order, we first need to enumerate all the possible Riemann scalars since the explicit list is never mentioned in the literature. We employ the following inefficient but straightforward way: first enumerate all possible scalars that can be formed by the contraction of their indices, taking into the account the symmetries, but not the cyclic identity. Then for each resulting scalar we replace each of its Riemann tensor factor respectively with the cyclic identity
\begin{equation}
    R_{abcd} + R_{acdb} + R_{adbc} = 0
\end{equation}
with each replacement two new terms are obtained. If both terms are contained in the Riemann scalar list then this term is equivalent to the already known scalars by the cyclic identity and should be removed.

As mentioned in the text, we get too many solutions at quintic order, so we are not going to present the full set of them here, which is included in the supplementary material. Instead, we only present a small portion of the dimension-generic solutions below.
\begin{subequations}\label{eq:quintic-qtgs}
    \begin{align}
        \mathcal Q^{(5), 1} & = -2 RR\indices{_a_b^e^f}R\indices{^a^b^c^d}R\indices{_c_e^g^h}R\indices{_d_f_g_h}+RR\indices{_a_b^e^f}R\indices{^a^b^c^d}R\indices{_c_d^g^h}R\indices{_e_f_g_h}\label{eq:quintic-qtg-1} \\
        \mathcal Q^{(5), 2} & = 2R\indices{^a^b}R\indices{^c^d}R\indices{_a_c^e^f}R\indices{_b_e^g^h}R\indices{_d_f_g_h}-2R\indices{^a^b}R\indices{^c^d}R\indices{_a^e_c^f}R\indices{_b_e^g^h}R\indices{_d_f_g_h}\nonumber\\
        &\quad+R\indices{_a^c}R\indices{^a^b}R\indices{_b^d^e^f}R\indices{_c_d^g^h}R\indices{_e_f_g_h} \label{eq:quintic-qtg-2}\\
        \mathcal Q^{(5), 3} & = \frac{8 (3 D-7) R^2}{3 D^2-15 D+16}R\indices{_a^c}R\indices{^a^b}R\indices{_b_c}-\frac{12 (D-1) R^3}{3 D^2-15 D+16}R\indices{_a_b}R\indices{^a^b}+\frac{4 R^5}{3 D^2-15 D+16}\nonumber\\
        & \quad + \frac{12 \left(D^2-5 D+8\right) R^2}{3 D^2-15 D+16}R\indices{^a^b}R\indices{^c^d}R\indices{_a_c_b_d}+\frac{6 (D-3) R^3}{3 D^2-15 D+16}R\indices{_a_b_c_d}R\indices{^a^b^c^d}\nonumber\\
        & \quad -\frac{12 (D-3) (D-1) R^2}{3 D^2-15 D+16}R\indices{^a^b}R\indices{_a^c^d^e}R\indices{_b_c_d_e}\nonumber\\
        &\quad-\frac{4 \left(D^3-9 D^2+26 D-22\right) R^2}{3 D^2-15 D+16}R\indices{_a^e_c^f}R\indices{^a^b^c^d}R\indices{_b_e_d_f}\nonumber\\
        & \quad +R^2R\indices{_a_b^e^f}R\indices{^a^b^c^d}R\indices{_c_d_e_f}\label{eq:quintic-qtg-3} \\
        \mathcal Q^{(5), 4} & = \frac{4 \left(D^3-5 D^2+8 D-8\right)}{D^4-11 D^3+44 D^2-72 D+36}R\indices{_a^c}R\indices{^a^b}R\indices{_b^d}R\indices{_c_d}R\nonumber\\
        & \quad -\frac{4 \left(D^4-10 D^3+39 D^2-68 D+40\right)}{(D-2) \left(D^4-11 D^3+44 D^2-72 D+36\right)}R\indices{_a_b}R\indices{^a^b}R\indices{_c_d}R\indices{^c^d}R\nonumber\\
        & \quad -\frac{16 \left(D^2-4 D+2\right)}{D^4-11 D^3+44 D^2-72 D+36}R\indices{_a^c}R\indices{^a^b}R\indices{_b_c}R^2\nonumber\\
        & \quad +\frac{2 \left(3 D^3-14 D^2+14 D-4\right)}{(D-2) \left(D^4-11 D^3+44 D^2-72 D+36\right)}R\indices{_a_b}R\indices{^a^b}R^3\nonumber\\
        & \quad -\frac{2 \left(D^2-4 D+2\right)}{(D-2) \left(D^4-11 D^3+44 D^2-72 D+36\right)}R^5\nonumber\\
        & \quad -\frac{8 \left(2 D^2-11 D+14\right)}{D^4-11 D^3+44 D^2-72 D+36}R\indices{^a^b}R\indices{^c^d}R^2R\indices{_a_c_b_d}\nonumber\\
        & \quad -\frac{D^2-6 D+10}{D^3-8 D^2+20 D-12}R^3R\indices{_a_b_c_d}R\indices{^a^b^c^d}+\frac{4 \left(D^2-5 D+6\right)}{D^3-8 D^2+20 D-12}R\indices{^a^b}R^2R\indices{_a^c^d^e}R\indices{_b_c_d_e}\nonumber\\
        & \quad +\frac{8 \left(D^3-5 D^2+4 D+4\right)}{D^4-11 D^3+44 D^2-72 D+36}R\indices{_a^c}R\indices{^a^b}R\indices{^d^e}RR\indices{_b_d_c_e}\nonumber\\
        & \quad +\frac{2 \left(D^4-10 D^3+35 D^2-46 D+16\right)}{D^4-11 D^3+44 D^2-72 D+36}R\indices{^a^b}R\indices{^c^d}RR\indices{_a_c^e^f}R\indices{_b_d_e_f}\nonumber\\
        & \quad -\frac{4 \left(D^4-9 D^3+28 D^2-32 D+8\right)}{D^4-11 D^3+44 D^2-72 D+36}R\indices{^a^b}R\indices{^c^d}RR\indices{_a^e_c^f}R\indices{_b_e_d_f}+R\indices{_a_b}R\indices{^a^b}RR\indices{_c_d_e_f}R\indices{^c^d^e^f}\label{eq:quintic-qtg-4}\\
        \mathcal Q^{(5), 5} & = \frac{2 \left(D^5-14 D^4+67 D^3-142 D^2+132 D-32\right)}{(D-4) (D-3) (D-2) \left(D^3-8 D^2+20 D-12\right)}R\indices{_a^c}R\indices{^a^b}R\indices{_b^d}R\indices{_c_d}R\nonumber\\
        &\quad +\frac{2 \left(5 D^4-45 D^3+154 D^2-236 D+128\right)}{(D-4) (D-3) (D-2)^2 \left(D^3-8 D^2+20 D-12\right)}R\indices{_a_b}R\indices{^a^b}R\indices{_c_d}R\indices{^c^d}R\nonumber\\
        &\quad -\frac{D^5-13 D^4+50 D^3-80 D^2+68 D-8}{(D-4) (D-3) (D-2) \left(D^3-8 D^2+20 D-12\right)}R\indices{_a^c}R\indices{^a^b}R\indices{_b_c}R^2\nonumber\\
        &\quad -\frac{-D^5+19 D^4-100 D^3+222 D^2-232 D+88}{(D-4) (D-3) (D-2)^2 \left(D^3-8 D^2+20 D-12\right)}R\indices{_a_b}R\indices{^a^b}R^3\nonumber\\
        &\quad +\frac{2 \left(D^3-5 D^2+7 D-4\right)}{(D-4) (D-3) (D-2)^2 \left(D^3-8 D^2+20 D-12\right)}R^5\nonumber\\
        &\quad -\frac{2 \left(D^4-13 D^3+56 D^2-106 D+88\right)}{(D-4) (D-3) \left(D^3-8 D^2+20 D-12\right)}R\indices{^a^b}R\indices{^c^d}R^2R\indices{_a_c_b_d}\nonumber\\
        &\quad -\frac{D^2-10 D+22}{2 (D-4) \left(D^3-8 D^2+20 D-12\right)}R^3R\indices{_a_b_c_d}R\indices{^a^b^c^d}\nonumber\\
        &\quad -\frac{-3 D^3+28 D^2-64 D+12}{2 (D-4) \left(D^3-8 D^2+20 D-12\right)}R\indices{^a^b}R^2R\indices{_a^c^d^e}R\indices{_b_c_d_e}\nonumber\\
        &\quad +\frac{2 \left(D^5-13 D^4+60 D^3-126 D^2+116 D-16\right)}{(D-4) (D-3) \left(D^3-8 D^2+20 D-12\right)}R\indices{_a^c}R\indices{^a^b}R\indices{^d^e}RR\indices{_b_d_c_e}\nonumber\\
        &\quad -\frac{-3 D^4+34 D^3-141 D^2+238 D-112}{(D-4) (D-3) \left(D^3-8 D^2+20 D-12\right)}R\indices{^a^b}R\indices{^c^d}RR\indices{_a_c^e^f}R\indices{_b_d_e_f}\nonumber\\
        &\quad -\frac{2 \left(D^4-8 D^3+19 D^2-12 D+4\right)}{(D-3) \left(D^3-8 D^2+20 D-12\right)}R\indices{^a^b}R\indices{^c^d}RR\indices{_a^e_c^f}R\indices{_b_e_d_f}\nonumber\\
        &\quad -\frac{3-D}{D-4}R^2R\indices{_a^e_c^f}R\indices{^a^b^c^d}R\indices{_b_e_d_f}\nonumber\\
        &\quad -\frac{D^2-3 D}{D-4}R\indices{^a^b}RR\indices{_a^c^d^e}R\indices{_b^f_d^g}R\indices{_c_f_e_g}\nonumber\\
        &\quad +R\indices{^a^b}RR\indices{_a^c_b^d}R\indices{_c^e^f^g}R\indices{_d_e_f_g}\label{eq:quintic-qtg-5}\\
        \mathcal Q^{(5), 6} & = -16 (2 D^9-57 D^8+639 D^7-3863 D^6+14059 D^5-31812 D^4+43724 D^3\nonumber\\
        &\quad -32916 D^2+9424 D+1088)/[(D-4) (D-3) (D-2) \left(D^3-8 D^2+20 D-12\right)\nonumber\\
        &\quad \left(D^5-14 D^4+79 D^3-224 D^2+316 D-170\right)]R\indices{_a^c}R\indices{^a^b}R\indices{_b^d}R\indices{_c_d}R\nonumber\\
        & \quad -16 (D^{10}-23 D^9+262 D^8-1914 D^7+9661 D^6-34319 D^5+85300 D^4-144708 D^3\nonumber\\
        &\quad+159028 D^2-101552 D+28480)/[(D-4) (D-3) (D-2)^2 \left(D^3-8 D^2+20 D-12\right)\nonumber\\
        &\quad \left(D^5-14 D^4+79 D^3-224 D^2+316 D-170\right)]R\indices{_a_b}R\indices{^a^b}R\indices{_c_d}R\indices{^c^d}R\nonumber\\
        & \quad + 32 (3 D^{11}-99 D^{10}+1368 D^9-10752 D^8+54161 D^7-184897 D^6+437712 D^5\nonumber\\
        &\quad-717804 D^4+794228 D^3-556336 D^2+214928 D-32032) / [(D-4) (D-3) (D-2)\nonumber\\
        &\quad\left(3 D^2-15 D+16\right) \left(D^3-8 D^2+20 D-12\right) (D^5-14 D^4+79 D^3-224 D^2\nonumber\\
        &\quad +316 D-170)]R\indices{_a^c}R\indices{^a^b}R\indices{_b_c}R^2\nonumber\\
        & \quad + 32 (3 D^{11}-45 D^{10}+183 D^9+869 D^8-12856 D^7+66448 D^6-201530 D^5\nonumber\\
        &\quad+395034 D^4-507818 D^3+413656 D^2-193024 D+39008) / [(D-4) (D-3)\nonumber\\
        &\quad(D-2)^2 \left(3 D^2-15 D+16\right) \left(D^3-8 D^2+20 D-12\right)\nonumber\\
        &\quad\left(D^5-14 D^4+79 D^3-224 D^2+316 D-170\right)]R\indices{_a_b}R\indices{^a^b}R^3\nonumber\\
        & \quad -4 (17 D^{10}-356 D^9+3319 D^8-18114 D^7+63896 D^6-151338 D^5+240888 D^4\nonumber\\
        &\quad-247784 D^3+147192 D^2-35360 D-2816)/[(D-4) (D-3) (D-2)^2 \nonumber\\
        &\quad\left(3 D^2-15 D+16\right) \left(D^3-8 D^2+20 D-12\right) (D^5-14 D^4+79 D^3-224 D^2\nonumber\\
        &\quad+316 D-170)]R^5\nonumber\\
        & \quad + 32 (3 D^{10}-99 D^9+1361 D^8-10615 D^7+52980 D^6-178926 D^5+417676 D^4\nonumber\\
        &\quad-669720 D^3+708764 D^2-447392 D+127552)/[(D-4) (D-3) \left(3 D^2-15 D+16\right) \nonumber\\
        &\quad\left(D^3-8 D^2+20 D-12\right) \left(D^5-14 D^4+79 D^3-224 D^2+316 D-170\right)]R\indices{^a^b}R\indices{^c^d}R^2R\indices{_a_c_b_d}\nonumber\\
        & \quad -4 (3 D^9-78 D^8+943 D^7-6838 D^6+32308 D^5-102114 D^4+214432 D^3\nonumber\\
        &\quad -286992 D^2+221088 D-74336)/[(D-4) \left(3 D^2-15 D+16\right) \nonumber\\
        &\quad \left(D^3-8 D^2+20 D-12\right) \left(D^5-14 D^4+79 D^3-224 D^2+316 D-170\right)]R^3R\indices{_a_b_c_d}R\indices{^a^b^c^d}\nonumber\\
        & \quad - 16 (3 D^9-84 D^8+948 D^7-5760 D^6+20693 D^5-44340 D^4+52004 D^3\nonumber\\
        &\quad-21480 D^2-14272 D+13008)/[(D-4) \left(3 D^2-15 D+16\right) \left(D^3-8 D^2+20 D-12\right)\nonumber\\
        &\quad\left(D^5-14 D^4+79 D^3-224 D^2+316 D-170\right)]R\indices{^a^b}R^2R\indices{_a^c^d^e}R\indices{_b_c_d_e}\nonumber\\
        & \quad -64 (D^9-24 D^8+245 D^7-1405 D^6+4982 D^5-11217 D^4+15640 D^3\nonumber\\
        &\quad-12198 D^2+3768 D+352)/[D-4) (D-3) \left(D^3-8 D^2+20 D-12\right) (D^5\nonumber\\
        &\quad-14 D^4+79 D^3-224 D^2+316 D-170)]R\indices{_a^c}R\indices{^a^b}R\indices{^d^e}RR\indices{_b_d_c_e}\nonumber\\
        & \quad +16 (D^9-28 D^8+336 D^7-2286 D^6+9737 D^5-26860 D^4+47638 D^3\nonumber\\
        &\quad-51642 D^2+30272 D-7024)/[(D-4) (D-3) \left(D^3-8 D^2+20 D-12\right) \nonumber\\
        &\quad\left(D^5-14 D^4+79 D^3-224 D^2+316 D-170\right)]R\indices{^a^b}R\indices{^c^d}RR\indices{_a_c^e^f}R\indices{_b_d_e_f}\nonumber\\
        & \quad +\frac{32 \left(D^8-18 D^7+134 D^6-532 D^5+1203 D^4-1534 D^3+1082 D^2-544 D+256\right)}{(D-3) \left(D^3-8 D^2+20 D-12\right) \left(D^5-14 D^4+79 D^3-224 D^2+316 D-170\right)}\nonumber\\
        &\quad R\indices{^a^b}R\indices{^c^d}RR\indices{_a^e_c^f}R\indices{_b_e_d_f}\nonumber\\
        & \quad -\frac{32 \left(2 D^7-35 D^6+263 D^5-1098 D^4+2743 D^3-4087 D^2+3352 D-1164\right)}{(D-4) \left(3 D^2-15 D+16\right) \left(D^5-14 D^4+79 D^3-224 D^2+316 D-170\right)}\nonumber\\
        &\quad R^2R\indices{_a^e_c^f}R\indices{^a^b^c^d}R\indices{_b_e_d_f}\nonumber\\
        & \quad +\frac{32 \left(D^6-14 D^5+82 D^4-254 D^3+433 D^2-380 D+132\right)}{(D-4) \left(D^5-14 D^4+79 D^3-224 D^2+316 D-170\right)}R\indices{^a^b}RR\indices{_a^c^d^e}R\indices{_b^f_d^g}R\indices{_c_f_e_g}\nonumber\\
        & \quad -\frac{2 \left(D^5-10 D^4+39 D^3-74 D^2+68 D-24\right)}{D^5-14 D^4+79 D^3-224 D^2+316 D-170}RR\indices{_a_b^e^f}R\indices{^a^b^c^d}R\indices{_c_e^g^h}R\indices{_d_f_g_h}\nonumber\\
        & \quad +RR\indices{_a_b_c_d}R\indices{^a^b^c^d}R\indices{_e_f_g_h}R\indices{^e^f^g^h}\label{eq:quintic-qtg-6}
    \end{align}
\end{subequations}

\section{Results of holographic shear viscosity}
In this section we present the values of the coefficients $a$, $b$ in \eqref{eq:sv-contrib-form} for dimension-generic quartic and quintic \ac{sqt} terms. The coefficients for quartic case are given in table \ref{tab:quartic-sv-coefs} below.
\begin{longtable}{|c|l|}
    \hline
    $a_{1}$ & $-\frac{(D-4) \left(2 D^8-29 D^7+145 D^6-245 D^5-101 D^4+226 D^3+1362 D^2-1984 D+304\right)}{16 (D-2) (D-1) D \left(D^3-9 D^2+26 D-22\right)}$ \\$b_{1}$ & $\frac{(D-4) \left(2 D^7-33 D^6+188 D^5-375 D^4-224 D^3+1722 D^2-1744 D+304\right)}{16 (D-2) (D-1) D \left(D^3-9 D^2+26 D-22\right)}$\\
    \hline
    $a_{2}$ & $-\frac{(D-4) \left(D^8-15 D^7+98 D^6-344 D^5+573 D^4+9 D^3-1376 D^2+1366 D-152\right)}{8 (D-2) (D-1) D \left(D^3-9 D^2+26 D-22\right)}$ \\$b_{2}$ & $\frac{(D-4) \left(D^7-15 D^6+108 D^5-481 D^4+1303 D^3-1930 D^2+1246 D-152\right)}{8 (D-2) (D-1) D \left(D^3-9 D^2+26 D-22\right)}$\\
    \hline
    $a_{3}$ & $-\frac{(D-4) \left(D^8-13 D^7+45 D^6+51 D^5-466 D^4+178 D^3+1828 D^2-2248 D+304\right)}{16 (D-2) (D-1) D \left(D^3-9 D^2+26 D-22\right)}$ \\$b_{3}$ & $\frac{(D-4) \left(D^7-16 D^6+71 D^5+38 D^4-1002 D^3+2452 D^2-2008 D+304\right)}{16 (D-2) (D-1) D \left(D^3-9 D^2+26 D-22\right)}$\\
    \hline
    $a_{4}$ & $-\frac{(D-4) \left(D^5-8 D^4+17 D^3+3 D^2-37 D+4\right)}{2 (D-2) (D-1) D}$ \\$b_{4}$ & $\frac{(D-4) \left(D^4-10 D^3+32 D^2-37 D+4\right)}{2 (D-2) (D-1) D}$\\
    \hline
    $a_{5}$ & $-\frac{(D-4) \left(D^5-8 D^4+17 D^3+3 D^2-37 D+4\right)}{(D-2) (D-1) D}$ \\$b_{5}$ & $\frac{(D-4) \left(D^4-10 D^3+32 D^2-37 D+4\right)}{(D-2) (D-1) D}$\\
    \hline
    $a_{6}$ & $-\frac{(D-4)^2 (D-3) (D+1)}{4 (D-2)}$ \\$b_{6}$ & $\frac{(D-4)^2 (D-3)}{4 (D-2)}$\\
    \hline
    $a_{7}$ & $\frac{(D-4) (D-2)^2 \left(4 D^3-9 D^2-55 D+8\right)}{32 D \left(D^3-9 D^2+26 D-22\right)}$ \\$b_{7}$ & $-\frac{(D-4) (D-2)^2 \left(D^3+3 D^2-40 D+8\right)}{32 D \left(D^3-9 D^2+26 D-22\right)}$\\
    \hline
    $a_{8}$ & $\frac{(D-4) (D-2)^2}{16 D}$ \\$b_{8}$ & $-\frac{(D-4) (D-2)^2}{16 D}$\\
    \hline
    $a_{9}$ & $\frac{3 (D-5) (D-4) (D-2)^2 \left(D^2+5 D-2\right)}{32 \left(D^3-9 D^2+26 D-22\right)}$ \\$b_{9}$ & $-\frac{3 (D-5) (D-4) (D-2)^2 (2 D-1)}{16 \left(D^3-9 D^2+26 D-22\right)}$\\
    \hline
    \caption{Coefficients $a$, $b$ of the shear viscosity contribution from quartic \ac{sqt} terms. Each entry corresponds to the solution with the same index in table \ref{tab:quartic-qtg-dimgen-sol}. Entries with zero coefficients are omitted.}
    \label{tab:quartic-sv-coefs}
\end{longtable}
For quintic \ac{sqt} terms, we only present the coefficients of the terms presented in the previous section in \eqref{eq:quintic-qtgs}.
\begin{longtable}{|c|l|}
    \hline
    $a_{3}$ & $\frac{3 (D-5) (D-4) (D-2)^2 D \left(D^2+8 D-4\right)}{16 \left(3 D^2-15 D+16\right)}$ \\$b_{3}$ & $-\frac{3 (D-5) (D-4) (D-2)^2 D (7 D-4)}{16 \left(3 D^2-15 D+16\right)}$\\
    \hline
    $a_{5}$ & $\frac{1}{64} (D-3) (D-2)^2 D$ \\$b_{5}$ & $-\frac{1}{64} (D-3) (D-2)^2 D$\\
    \hline
    $a_{6}$ & $-\frac{(D-3) (D-2)^2 \left(6 D^8-81 D^7+380 D^6-508 D^5-1660 D^4+7593 D^3-12210 D^2+8512 D-1792\right)}{4 \left(3 D^2-15 D+16\right) \left(D^5-14 D^4+79 D^3-224 D^2+316 D-170\right)}$ \\$b_{6}$ & $\frac{(D-3) (D-2)^2 \left(21 D^7-336 D^6+2194 D^5-7582 D^4+15001 D^3-16834 D^2+9472 D-1792\right)}{4 \left(3 D^2-15 D+16\right) \left(D^5-14 D^4+79 D^3-224 D^2+316 D-170\right)}$\\
    \hline
    \caption{Coefficients $a$, $b$ of the shear viscosity contribution from quintic \ac{sqt} terms. Each entry corresponds to the solution with the same indices in \eqref{eq:quintic-qtgs}. Entries with vanishing coefficients are omitted.}
    \label{tab:quintic-sv-coefs}
\end{longtable}

\acknowledgments
I thank Yi Ling and Hong Lü for helpful discussions, I am also grateful to Ying Chen for supporting my own study projects.

\bibliography{refs}

\end{document}